\documentclass{sig-alternate}
\usepackage{color}
\usepackage{subfigure}
\usepackage{graphicx}
\usepackage{multirow}
\usepackage{times}

\enlargethispage{\baselineskip} 

\newcommand{\denselistA}{ \itemsep -0.45pt\topsep-5pt\partopsep-7pt }

\begin{document}
\CopyrightYear{2011}
\crdata{978-1-4503-0261-6/11/06}
\conferenceinfo{EC}{'11, June 5-9, 2011, San Jose, California, USA.}
\clubpenalty=10000 
\widowpenalty = 10000

\newcommand{\hide}[1]{}
\newcommand{\semihide}[1]{{\tiny #1}}
\newcommand{\rev}[1]{\textsf{{\textcolor{red}{[[#1]]}}}}
\newcommand{\xhdr}[1]{\vspace{1.7mm}\noindent{{\bf #1.}}}

\title{The Role of Social Networks in Online Shopping:\\
Information Passing, Price of Trust, and Consumer Choice}

\numberofauthors{3}
\author{
\alignauthor
Stephen Guo\\
       \affaddr{Stanford University}\\
       \affaddr{Stanford, CA, USA}\\
       \email{sdguo@cs.stanford.edu}
\alignauthor
Mengqiu Wang\\
       \affaddr{Stanford University}\\
       \affaddr{Stanford, CA, USA}\\
       \email{mengqiu@cs.stanford.edu}
\alignauthor
Jure Leskovec\\
       \affaddr{Stanford University}\\
       \affaddr{Stanford, CA, USA}\\
       \email{jure@cs.stanford.edu}
}

\maketitle

\begin{abstract}
While social interactions are critical to understanding consumer behavior, the
relationship between social and commerce networks has not been explored on a
large scale. We analyze Taobao, a Chinese consumer marketplace that is the
world's largest e-commerce website. What sets Taobao apart from its competitors
is its integrated instant messaging tool, which buyers can use to ask sellers
about products or ask other buyers for advice. In our study, we focus on how an individual's commercial transactions
are embedded in their social graphs. By studying triads and the directed
closure process, we quantify the presence of information passing and
gain insights into when different types of links form in the network.

Using seller ratings and review information, we then quantify a price of
trust. How much will a consumer pay for transaction with a trusted seller? We conclude by modeling
this consumer choice problem: if a buyer wishes to purchase a particular product, how does
(s)he decide which store to purchase it from? By analyzing the performance of
various feature sets in an information retrieval setting, we demonstrate how
the social graph factors into understanding consumer behavior.
\end{abstract}

\vspace{2mm} \noindent {\bf Categories and Subject Descriptors:}
H.2.8 {Database Management}: {Database Applications -- Data mining}

\vspace{1mm} \noindent {\bf General Terms:} Measurement;
Experimentation.

\vspace{1mm} \noindent {\bf Keywords:} E-Commerce, Viral Marketing,
Recommender Systems, Triadic Closure, Price of Trust.

\section{Introduction}
\label{sec:intro}
Use of personal social networks to gather information is fundamental to
purchasing behavior~\cite{dimaggio1998socially}. It is something so common in
our daily routine that we usually do not even make a note of it. When we make a
purchase from a retail store, we often speak beforehand to the shopkeeper
about suitable products. When we need to purchase something we are unfamiliar
with, we consult our friends and family for advice. When we purchase a popular
new product, we have an urge to tell everyone we know about it.

Although personal social networks are implicit in the offline shopping
experience, their introduction to the online world is a relatively new
phenomenon. E-commerce websites, such as Amazon, eBay and Epinions, have successfully integrated
product reviews, recommendations, search and product comparison, but they have been
much slower at adopting social networking features as a part of customer
experience. Recommendation engines and product comparison sites help consumers discover new products and receive
more accurate evaluations, however they cannot completely substitute for the personalized
recommendations and information that one receives from a friend or relative. 
Basic behavioral psychology drives consumers to value and trust their
friends' purchasing decisions more than anonymous opinions. For example, a Lucid Marketing survey found that
68\% of individuals consulted friends and relatives before purchasing home
electronics~\cite{burke03lucid}.

Understanding how social networks are used and how they shape purchasing
decisions is one of the fundamental interests of e-commerce. Only recently have social networks been used in e-commerce applications to some
success. For example, group purchasing companies such as Groupon and LivingSocial allow
consumers to come together to buy products in bulk and save money, while social
shopping sites such as Kaboodle provide consumers the ability to share shopping
lists with each other. The use of social networks in online shopping provides
marketers and businesses with new revenue opportunities, while providing
consumers with product information and both economic and social rewards for sharing~\cite{hennig2004electronic}.

\xhdr{Present work} When discussing the relationship between electronic
commerce and social networks, various questions come to mind. How do friends influence
consumer purchasing decisions and product adoption? What factors influence the
success of word-of-mouth product recommendations?  How does social influence
and reputation affect commercial activity? In this paper, we will address these
questions through a detailed study of the world's largest e-commerce website Taobao.

The fundamental process we focus upon throughout this study is what we
term \textit{information passing}: an individual
purchases a product, then messages a friend, what is the likelihood that the
friend will then purchase the product? Where will he purchase it from? Understanding the flow of social
influence in commerce networks is an important question.
For example, information passing provides insight into how companies can
structure online viral marketing campaigns to target consumers. It can also be used to optimize algorithms within product recommendation engines.
However important information passing is to electronic commerce, it still has not been well studied on a large scale due to the inaccessibility of suitable data.

To facilitate our research, we obtained a dataset describing the behavior of one million
users in the world's largest e-commerce network Taobao. 
Taobao connects buyers and sellers, and provides an integrated instant messaging platform for communication
among all its users. By modeling Taobao as a network of three types of edges
(trades, messages, contacts), we are able to directly study how social communication and commercial transactions are
interrelated in an online setting. Our study provides insights into three main aspects of the
social-commercial relationship: information passing, the price of trust, and consumer choice prediction.

We begin our study of social commerce by quantifying the presence of information passing through analysis of
triadic closure processes. We show that the influence of information passing is directly
proportional to message strength, and is inversely proportional to product
price, as well as the time between the purchase and the recommendation. Additionally,
we explain how information passing varies greatly between different product
categories. We then investigate the general edge formation process in the
context of directed triadic closure, and demonstrate that the formation of
triad-closing message and trade edges is highly dependent upon user roles (buyer or seller). Our results indicate that
information passing via buyer-buyer communication is one of the primary drivers
of purchasing.

A subtle point regarding information passing is that the spread of product
recommendations, through word-of-mouth, inherently relies upon a notion of
buyer-buyer \textit{trust}. Trust, from the perspective of social psychology, can be
defined as perceived credibility or benevolence to the
target~\cite{doney1997examination}. 
In the context of electronic marketplaces,
buyer-seller trust, most directly encapsulated by seller reputations and ratings, are the
natural concept to study. A fundamental idea behind the nature of
trust is its price. How much extra will a buyer pay for transaction security with a highly-rated
seller? Although an intuitive idea, initial studies did not find evidence for a price of
trust~\cite{resnick2002trust}, and only recently has a price of trust been established in small, controlled, and
single product settings~\cite{resnick2006value,houser2006reputation,lucking2007pennies}.
In our study, we analyze transaction information across over 10,000 products.
Using the overall customer satisfaction (i.e., average rating) of the seller,
we observe a small but super-linear effect of the seller rating upon the price
premium they can charge and still engage in transactions.

To further study the relationship between social networks and consumer
behavior, we then consider the question, ``How does an online
consumer decide upon a seller to purchase from when there are many sellers offering the same product?'' 
We model this question of \textit{consumer choice} through a machine learning
 task and predict which particular seller a buyer will purchase from, given that the sellers all offer the same relevant product.
Utilizing primarily social networking features, we construct
a model that can predict, for the case of a buyer choosing from among 10
possible sellers, the correct seller 42\% of the time, approximately 4 times
better than baseline. We also contrast a variety of feature sets (both
graph-based and product/seller metadata), and demonstrate that the social graph
is the most important feature in predicting consumer choice. In particular, the social graph is
a far better determinant of consumer behavior than metadata features such as
seller reputation or product price. Our results nicely connect to Granovetter's work, which argues that economic
transactions are embedded in dynamic social networks, and that an individual's social
graph dictates how they choose sellers to transact with~\cite{granovetter1985economic}. 

\xhdr{Further Related Work}
\label{sec:related}
For all of the importance of social networks in consumer shopping, though,
their use in electronic commerce still is not well understood. Previous
research examined the use of social networks in e-commerce, but has mostly
focused upon one aspect of the use of social networks, such as product
recommendations~\cite{hill2006network,leskovec2007dynamics,bhatt2010predicting}, product
recommendation engines~\cite{schafer1999recommender,sarwar2000analysis}, or
have been based upon a limited set of data~\cite{chevalier2006effect,iyengar2009friends}.

\vspace{1mm}
The electronic marketplace eBay is perhaps the most well studied e-commerce
website. Various aspects of eBay including auction efficiency~\cite{hu2008online}, product
recommendations~\cite{zheng2009substitutes}, seller
strategies~\cite{duong2010modeling}, and summarization~\cite{lu2009rated}, have
been studied. Closely related to our work on consumer choice prediction, Wu and Bolivar
created a model to predict item purchase probability~\cite{wu2009predicting}.
The primary difference here is that we utilize the
social networks of the buyers and sellers, along with product and user
metadata, to perform consumer choice prediction, whereas they consider only seller and product information.

Our study of information passing and triadic closure builds upon classical
works by Rapoport~\cite{rapoport1953spread} and
Granovetter~\cite{granovetter1973strength}. Triadic closure has been explored
in various settings: community growth~\cite{backstrom2006group}, link
prediction~\cite{liben2007link}, signed networks~\cite{leskovec2010predicting},
and social~\cite{leskovec2008microscopic} and
information~\cite{romero2010directed} network evolution. In contrast, we
demonstrate the existence of implicit recommendation behavior in a network not
specifically designed for information passing.

The paper proceeds as follows. In section 2, we describe the Taobao
dataset. In section 3, we analyze dyadic relationships in the network. In
section 4, we provide a detailed analysis of information passing and directed
triadic closure processes. In section 5, we quantify a price for trust. And
last, in section 6, we model the consumer choice prediction problem.

\section{Taobao network}
\label{sec:prelim}
The data we use in this study comes from the Chinese website Taobao, 
one of the world's largest electronic marketplaces, with over 370 million
registered users at the end of 2010.
Although transactions on Taobao can be either business-to-consumer, business-to-business, or
consumer-to-consumer, the bulk of the products are sold by online storefronts
operated by small businesses or individuals. 
Perhaps the most unique aspect of Taobao is its integrated in-browser instant
messaging platform, which allows us to correlate users' communication patterns
and purchasing behavior. Any user can purchase goods from other users, add other users onto their contact list, and message other
users. Note that non-contacts can message each other as well. 

\begin{table}[t]
 \centering
  \begin{tabular}{l||r|r|r|r}
    \text{Network} & \text{Nodes} & \text{Edges} & \text{Avg Deg} & \text{Avg CCF}\\
    \hline
    \hline\text{Contact} & 663,346 & 3,208,043 & 9.67 & 0.0135\\
    \hline\text{Message} & 750,158 & 3,908,339 & 5.21 & 0.0194\\
    \hline\text{Trade} & 1,000,000 & 1,337,497 & 1.34 & 0.0086
    \end{tabular}
  \vspace{-2mm}
  \caption{Dataset statistics.}
  \label{table:basicStats}
  \vspace{-5mm}
\end{table}

Our data is composed of all activities of the set of the first one million users that
engaged in at least one commercial transaction during September
1 through October 28, 2009. For each of these users, we have all
information regarding their transactions with other users in the set, where a
transaction is specified by a product identifier, price, quantity, and timestamp. 
We also obtained the contact lists and timestamps of messages exchanged between these users
during the two month observation period. Note that we do not have the contents
of the messages exchanged.

We model the Taobao network as a multigraph composed of directed
trade edges, directed message edges, and undirected contact/friendship
edges.\footnote{The multigraph is modeled such that if there are multiple trades or messages from one user to another, we aggregate it
all into a single directed edge, with supplementary message and trade
information being associated with that edge. There can be up to 5 edges between
a pair of users (2 directed trades, 2 directed messages, 1 undirected contact).} 
Table~\ref{table:basicStats} shows the basic statistics of the Taobao network
when each edge type is viewed as a separate network. The
3,908,339 directed message edges are equivalent to 2,241,729 undirected message
edges, as messages are often reciprocated during discourse between a pair of
users. In contrast, the 1,337,497 directed trade edges are equivalent to 1,336,502
undirected trade edges, as purchases are almost never reciprocated. 

Throughout the rest of this paper, when we refer to a node as a ``buyer'' or
``seller,'' we are speaking about its role in a particular transaction. Thus,
we do not a priori label the nodes as buyers or sellers, but we use these terms
to aid explanation.
As a point of reference, 968,149 users make at least one purchase, 69,494 users make at
least one sale, and 37,643 users make both a purchase and sale during the observation
period. The products purchased and sold by these users are classified by Taobao into 82
categories. 

\section{Dyadic relationships}
\label{sec:dyads}
\begin{figure}[t]
\centering
\includegraphics[width=0.40\textwidth]{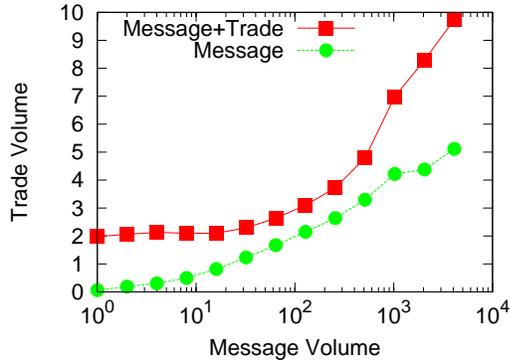}
  \vspace{-3mm}
\caption{Trade Volume versus Message Volume. \textit{Message} is computed over
all
pairs of users that exchange at least one message. \textit{Message+Trade} is
computed
over all pairs of users that exchange at least one message and one trade. }
\label{fig:mtEdge}
    \vspace{-3mm}
\end{figure}

To facilitate our goal of understanding how commercial transactions are embedded in the
social networks of buyers and sellers, we first examine dyadic relationships in Taobao. In particular, we are interested in
determining if messaging activity is correlated with trading
activity. We graph trade volume versus message volume
across pairs of users in Figure~\ref{fig:mtEdge}, and find that there is a positive increasing relationship between
message volume and trade volume. Ignoring all pairs of users that only message and do not trade, we see an even more
pronounced increasing relationship, displayed in the \textit{Message+Trade}
curve. The positive correlation between messaging and trading activity across dyads is
supportive of our hypothesis that commercial activity is embedded in social
networks. We will investigate the relationship between communication and
commerce in much greater detail in our study of triadic structures.

\begin{figure}[t]
\begin{minipage}[b]{0.23\textwidth}
\centering
\includegraphics[width=\textwidth]{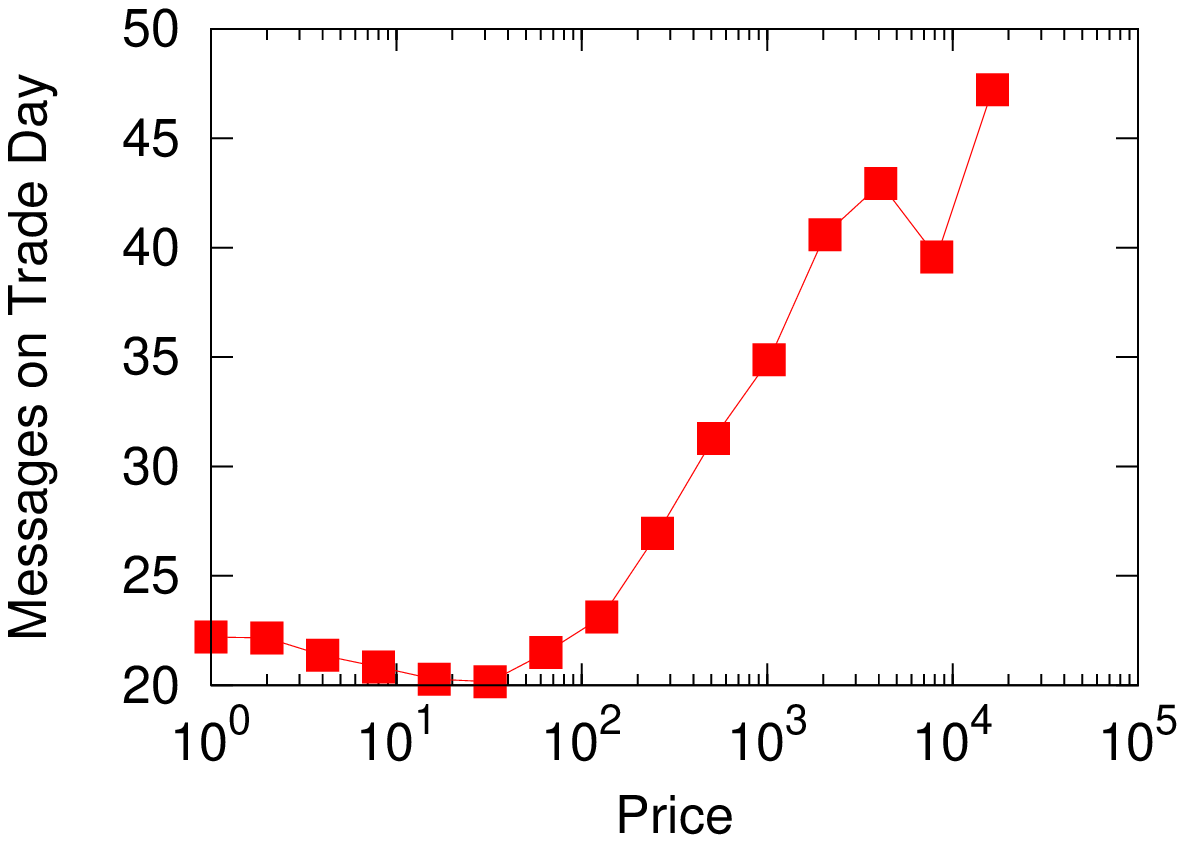}
\end{minipage}
\begin{minipage}[b]{0.23\textwidth}
\centering
\includegraphics[width=\textwidth]{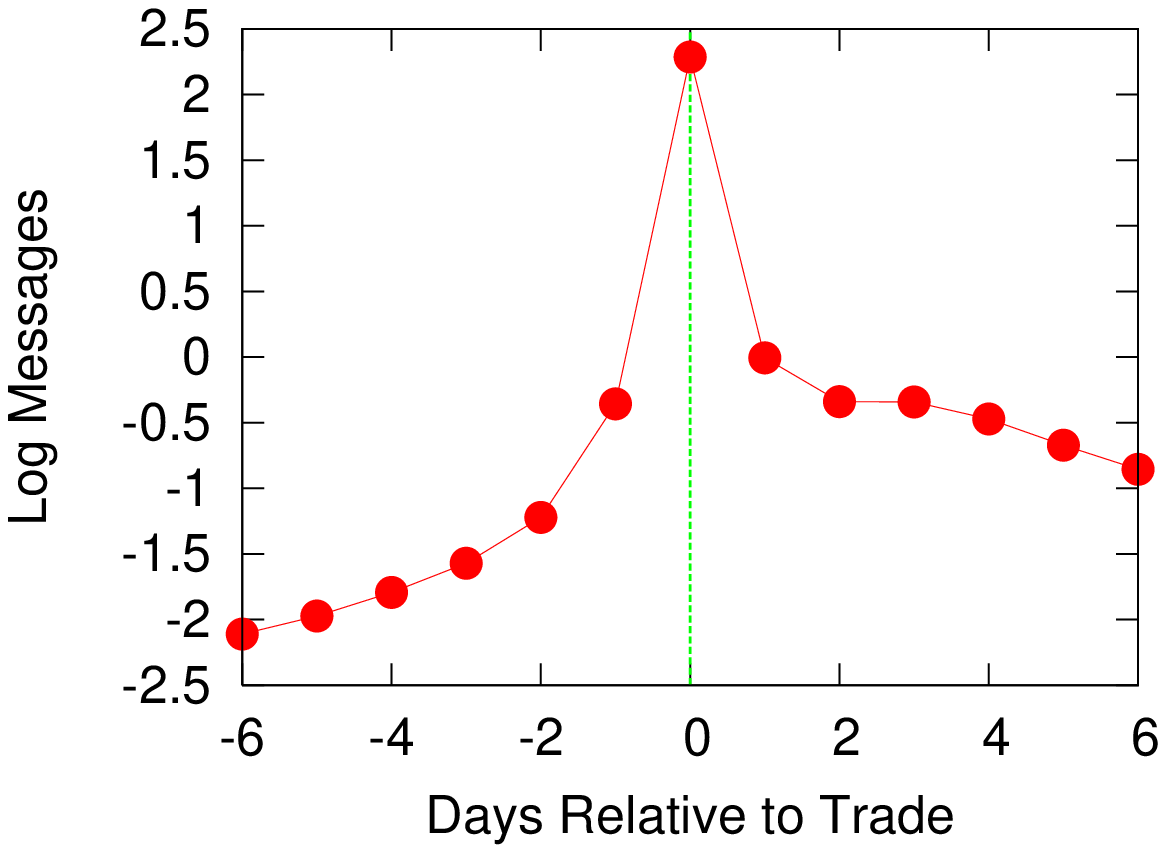}
\end{minipage}
  \vspace{-3mm}
\caption{Buyer-Seller Messages vs. Transaction Price (left), relative to Trade
  Date (right).}
\label{fig:buySellMsg}
    \vspace{-3mm}
\end{figure}

Focusing upon the subset of dyads between pairs of users who have
historically transacted, our question of interest is, ``Do buyers talk to
sellers more about expensive products?'' We expect that expensive
products are talked about more, but how much more are they talked about?
To answer this question, we count the number of messages sent from buyer to
seller on transaction date, assuming that at least one message is exchanged,
and plot it versus product price (in Chinese Yuan, CNY), displayed in
Figure~\ref{fig:buySellMsg}(left).
We find that the number of messages sent is relatively constant for
products of price below 100 CNY, then increases logarithmically for products of
higher price. This relationship can be explained by 
messaging being one of the primary tools in Taobao through which a buyer can minimize transaction risk.

In order to minimize transaction risk, one would expect that buyers speak to sellers often before
transaction to inquire about product details.
How often do buyers speak to sellers before and after trades?
We graph the number of buyer-seller messages versus trade date in Figure~\ref{fig:buySellMsg}(right). As expected, most
messages occur on the day of transaction, likely being product negotiation.
What we find particularly interesting is that post-trade messages are significantly more common than
pre-trade messages. In the Taobao system, buyers have an option of using an
escrow service, where the seller first ships the product, and payment is
exchanged after the buyer examines the product. The observed post-trade
messages are likely discussion regarding product satisfaction and payment confirmation.

\section{Information passing}
\label{sec:experiments}
From our study of dyadic buyer-seller relationships, we learn that messaging
activity is correlated with trading activity across pairs of users. However,
dyadic relationships are only the tip of the iceberg when thinking about
the interplay between communication and purchasing decisions in a social
commerce network. Imagine the following situation: a buyer
notices a deal offered at an electronic store, makes a purchase, then
messages his friend about the deal. Will the friend also make a purchase from
the same store? How large is the influence of the buyer?

\begin{figure}[t]
\begin{minipage}[b]{0.14\textwidth}
\centering
\includegraphics[width=\textwidth]{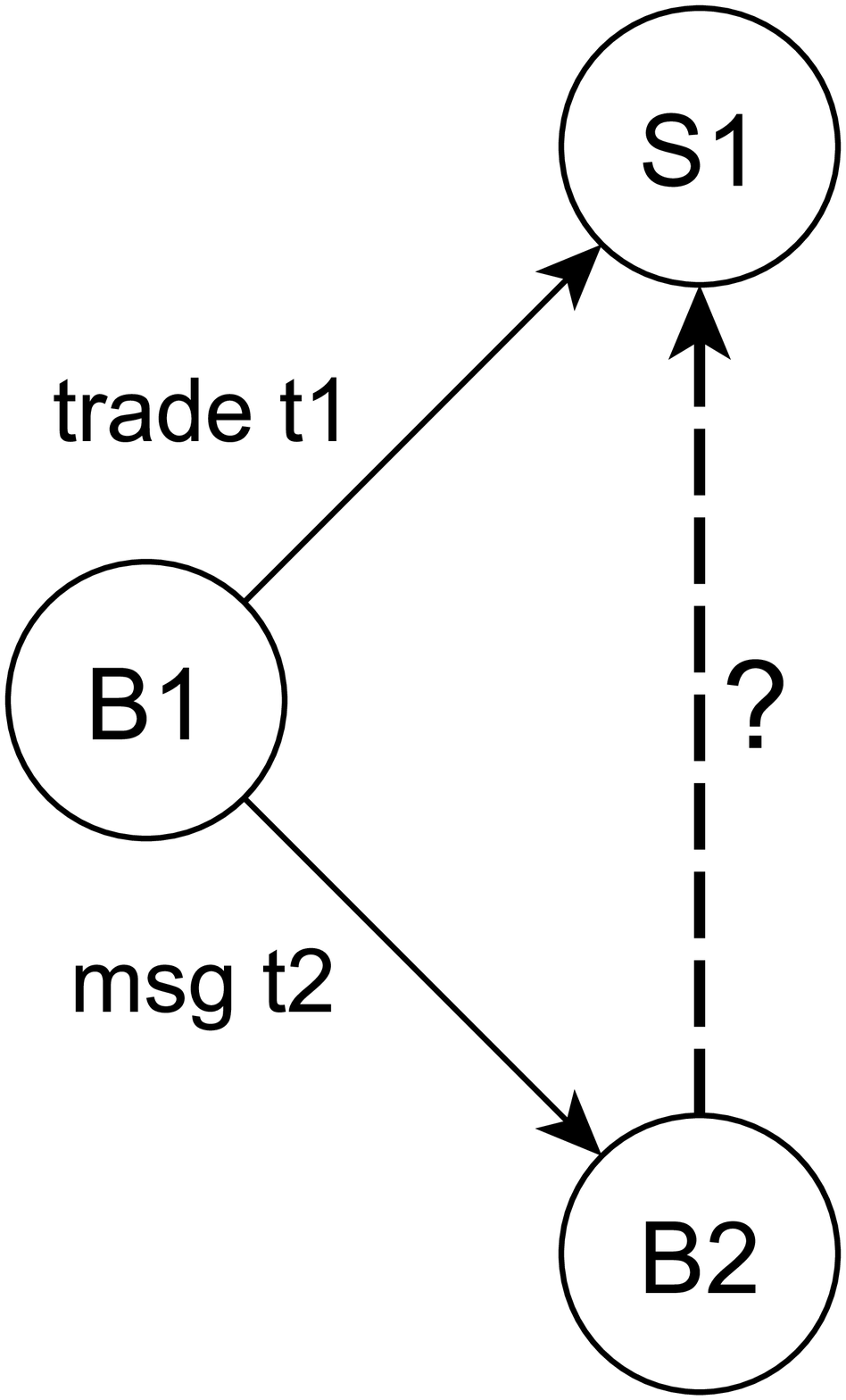}
\end{minipage}
\begin{minipage}[b]{0.3\textwidth}
\centering
\includegraphics[width=\textwidth]{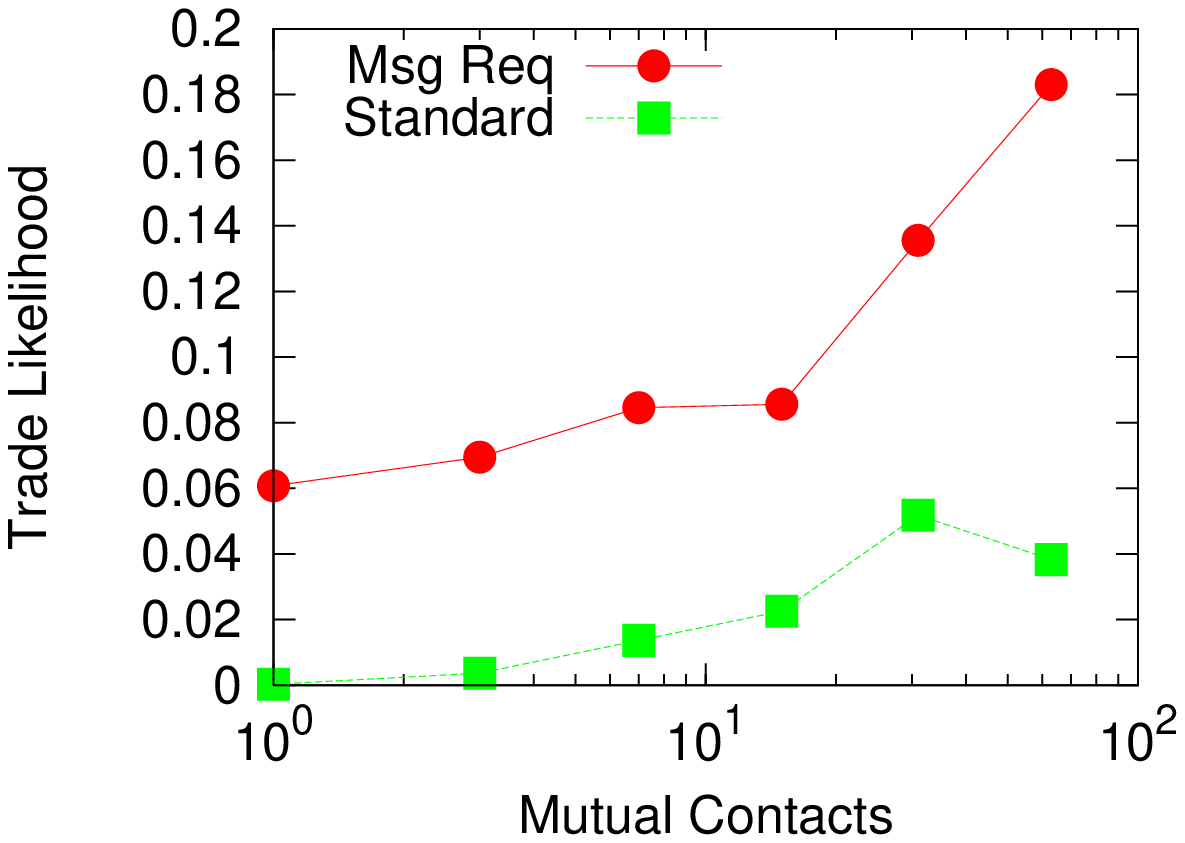}
\end{minipage}
  \vspace{-3mm}
\caption{Given that B1 first purchased from S1 and then talked to B2, will B2 purchase from S1? (left) and Mutual Contacts (right).}
\label{fig:mutual}
    \vspace{-3mm}
\end{figure}

We quantify and investigate this economic diffusion
behavior which we term \textit{information passing}, illustrated in
Figure~\ref{fig:mutual}(left). More formally, if buyer \textit{$B_{1}$} purchases from seller \textit{$S_{1}$} and then talks
to user \textit{$B_{2}$}, will user \textit{$B_{2}$} then purchase from seller
\textit{$S_{1}$} as well?\footnote{Again, we assign buyer and seller roles
with respect to particular transactions. So user \textit{$B_{2}$} can be a
seller and user \textit{$S_{1}$} can be a buyer in some other transactions.}
In the subsequent sections, we analyze information passing through the study of (1) local mutual neighborhoods in static networks, (2)
information passing in dynamic networks, (3) influences upon information passing, and (4) directed triadic closure.

\xhdr{Information Passing and Triadic Closure} We begin our study of
information passing by examining local neighborhoods in each edge type network
(trade, message, contact) of a static endtime snapshot of Taobao. We are interested in understanding how the mutual
relationships in one static network are correlated with the edge likelihood in another static network.
Of particular interest is the question of how \textit{social proximity} is
correlated with trade likelihood between a pair of users, where
social proximity is measured by the number of mutual contacts
between the pair. If we demonstrate that there is correlation between
social proximity and trade likelihood, then information
passing processes, as shown in Figure~\ref{fig:mutual}(left), are likely present in Taobao.

We find that the more mutual contacts a pair of users has, the greater the likelihood that they
engaged in a commercial transaction, labeled as \textit{Standard} in Figure~\ref{fig:mutual}(right).
If we restrict our attention to only users who have
exchanged at least one message (\textit{Msg Req}), then for a given number of
mutual contacts, the transaction probability is noticeably greater than
before.\footnote{To compare, the likelihood that a pair of direct contacts have
transacted historically is 0.089.}
From these results, we can infer that trades are more likely to be embedded in
the dense subgraphs of communication networks. This implies that social proximity and trade likelihood are correlated,
and are a signal that information
passing and product recommendation may be present in the Taobao network.
In general, a direct relationship in one network is not only embedded in
the local neighborhood of that relationship,
but is also positioned in the context of networks with other edge types. This suggests that edges in one network can be used to help
understand the link formation process in another. Building on this idea, we
next perform a similar experiment in a dynamic triadic closure setting.

\xhdr{Information Passing}
Following our examination of the static network, we study network relationships in the dynamic network.
In particular, we look at how the message and contact networks influence the
trade network by checking for the presence of information passing, as displayed in Figure~\ref{fig:mutual}(left).

To quantify information passing in the Taobao network, we measure the \textit{information passing success rate} of the network, which we
define as Prob(\textit{$B_{2}$} buys from \textit{$S_{1}$} at \textit{$t_{2} + \Delta$} |
\textit{$B_{1}$} buys from \textit{$S_{1}$} at \textit{$t_{1}$} and
\textit{$B_{1}$} messages \textit{$B_{2}$} at \textit{$t_{2}$},  \textit{$t_{2}$} >
\textit{$t_{1}$}).\footnote{We only consider the time \textit{$t_{2}$} corresponding to the first
message from \textit{$B_{1}$} to \textit{$B_{2}$} after \textit{$t_{1}$}. We
also add a requirement of $\Delta \leq$ 2 days, i.e., \textit{$B_{2}$} makes a purchase
soon after talking to \textit{$B_{1}$}, to dampen the effects of regular purchases that occur irrelevant to messaging behavior.}

Before computing the information passing success rate for the Taobao network,
we require a random baseline for comparison. For our baseline, we compute the information passing success rate of an edge-rewired
version of the Taobao network, where the edge-rewired network is constructed by
randomly rewiring all 3 types of edges in the original network, while leaving node degrees
and edge creation times unchanged.

We compute the information passing success rate over 3,906,354 node pair instances in the
original network and observe a probability of 0.00203. In contrast, the
information passing success rate of the rewired network is computed to be 0.00006.
The observed probability of recommendation success is two orders of
magnitude more likely than the random baseline, implying that information
passing in Taobao is statistically significant and non-random.

Having verified the presence of information passing by
checking for edge formation in the forward direction, we confirm the presence
of information passing in the reverse direction.
Suppose that \textit{$B_{1}$} buys from \textit{$S_{1}$} at time
\textit{$t_{1}$} and \textit{$B_{2}$} buys from the same \textit{$S_{1}$}
at time \textit{$t_{1}+\delta$}, we measure the number of messages exchanged
between \textit{$B_{1}$} and \textit{$B_{2}$} in the time
intervals \textit{Before} [$t_{1}-\delta$,$t_{1}$], \textit{Between}
[$t_{1}$,$t_{1}+\delta$],
and \textit{After} [$t_{1}+\delta$, $t_{1}+2\delta$] the purchases of
\textit{$B_{1}$} and \textit{$B_{2}$}.
One expects that if information passing is present, then most messages
exchanged betweeen the two buyers occur after \textit{$B_{1}$} purchases,
but before \textit{$B_{2}$} purchases.
Table~\ref{table:msgEdgeStudy} shows the messages for the 3
time periods versus $\delta$, averaged over all instances. We see that the largest proportion of messages exchanged between the buyers occur between
their corresponding trade dates. For example, when the buyers transact two days apart, twice as many messages are exchanged
\textit{Between} the purchase dates, as compared to \textit{Before} or \textit{After}. Since messages exchanged \textit{Between} are
more likely to be recommendations, this is additional evidence that information
passing is present in the Taobao network.

Through examination of both forward and backward processes, we demonstrate that
information passing is present in the Taobao network. In particular, we show
that the observed information passing success rate is two orders of magnitude
more likely than a random baseline. Prior studies providing evidence of information passing have been
primarily conducted in product recommendation networks~\cite{leskovec2007dynamics}. Our work shows that information
passing, involving both buyer-buyer and buyer-seller relationships, occurs implicitly in Taobao, where the communication tool is primarily intended
for buyer-seller communication. This result is significant because it illustrates how offline consumer behavior, asking or informing
a personal social network about products, is also manifested implicitly in online social commerce networks.

\begin{table}[t]
  \centering
  \begin{tabular}{c||r|r|r}
  \text{Days between purchases} & \text{Before} & \text{Between} & \text{After}\\
    \hline    \hline
    \text{1} & 4.16 & 5.29 & 4.76\\
    \text{2} & 7.78 & 14.29 & 7.76\\
    \text{3} & 8.60 & 10.52 & 7.44 \\
    \text{4} & 7.34 & 15.90 & 10.79\\
    \text{5} & 21.87 & 30.70 & 21.18
    \end{tabular}
      \vspace{-3mm}
  \caption{Messages between two buyers relative to their trade dates with the
    same seller.}
  \label{table:msgEdgeStudy}
  \vspace{-5mm}
\end{table}

\begin{figure*}[t]
\centering
\includegraphics[width=0.3\textwidth]{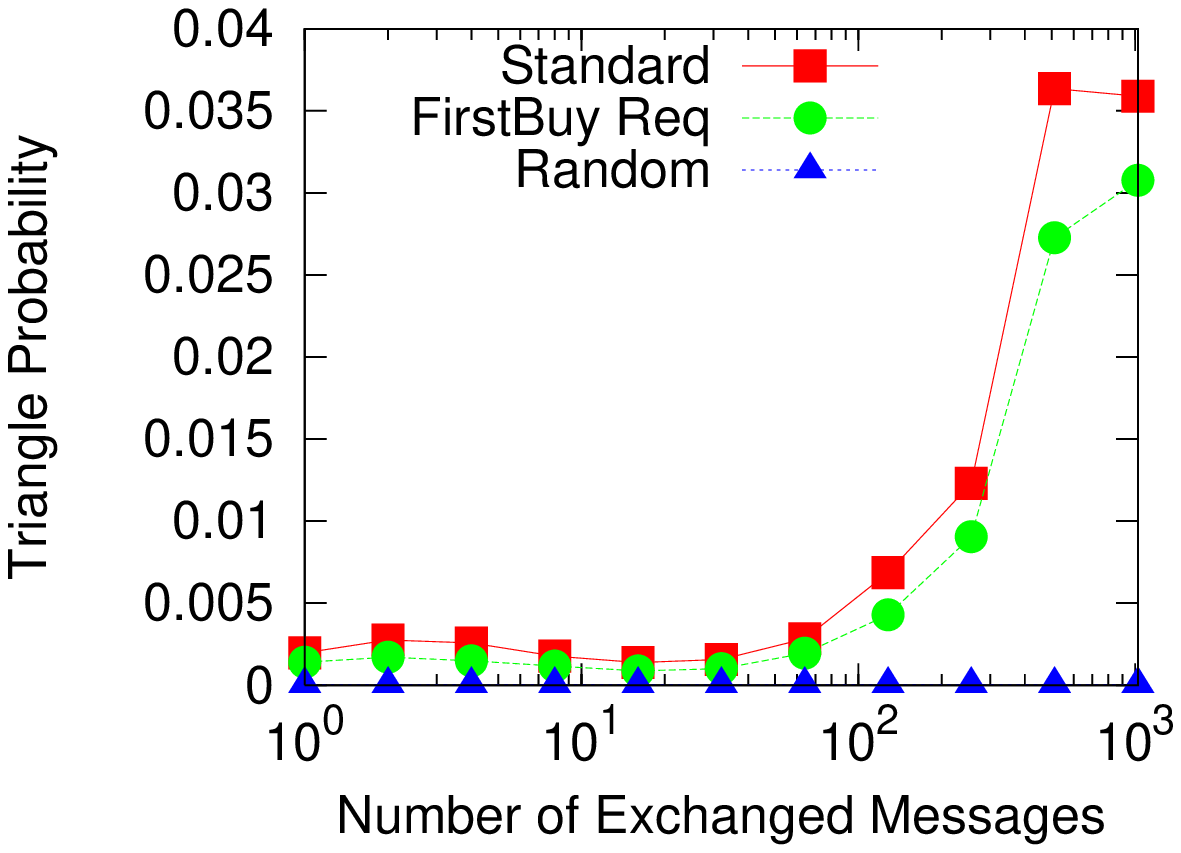}
\includegraphics[width=0.3\textwidth]{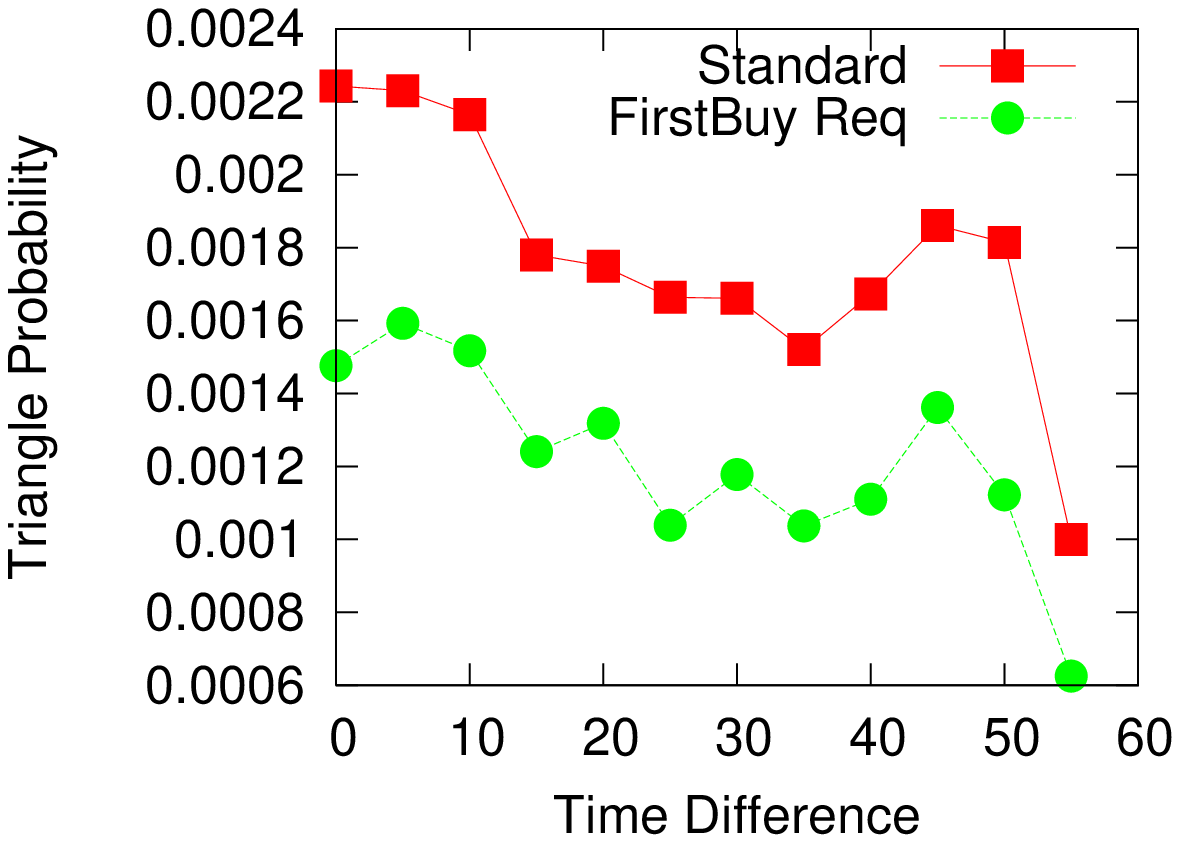}
\includegraphics[width=0.3\textwidth]{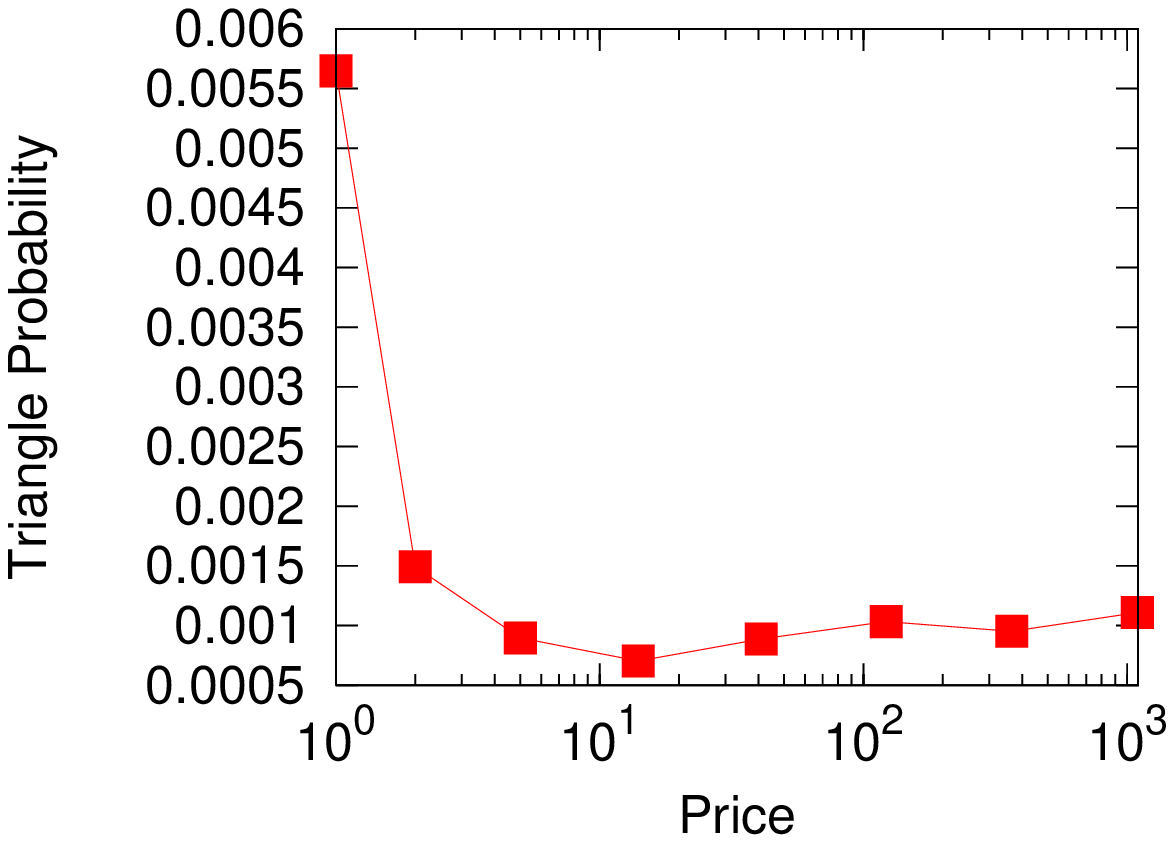}
  \vspace{-3mm}
\caption{Triadic Closure Probability given Message Strength (left), Time
Difference in Days (middle), and Price in CNY (right).}
\label{fig:closure}
    \vspace{-5mm}
\end{figure*}

\xhdr{Influences upon Information Passing}
Having demonstrated the existence of information passing in the Taobao network,
the next question is, ``What factors influence the success rate of
information passing?'' In the following experiments, we examine how information
passing success varies with respect to 4 variables: communication strength, time
difference, product price, and product category.

Perhaps the primary influence upon the success of information passing is the
 amount of communication between the initial buyer, \textit{$B_{1}$}, and his messaging
 partner, \textit{$B_{2}$}. One can hypothesize that the stronger the
 communication between the two users, the more likely that \textit{$B_{2}$}
 will also purchase the product from the seller that sold to \textit{$B_{1}$}.
Counting the number of messages exchanged within time window [\textit{$t_{1} - \delta$},
  \textit{$t_{1} + \delta$}], the \textit{Standard} curve of Figure~\ref{fig:closure}(left) shows the probability
of closure with $\delta$ = 3 days. As expected, the stronger the
communication between the two users, the more likely that \textit{$B_{2}$} will
be influenced by \textit{$B_{1}$}.
Adding a requirement to consider only users \textit{$B_{2}$} who have never purchased from
 \textit{$S_{1}$} historically (\textit{FirstBuy Req}), we get a slightly lower, but still
 similar likelihood curve.

Note that an alternative possible explanation for these findings is that
 both \textit{$B_{1}$} and \textit{$B_{2}$} are active users in Taobao
 who communicate with other users frequently, make more purchases, and are thus
 more likely to purchase from the same sellers. However, we demonstrate this hypothesis is incorrect by performing an experiment where we keep the communication network and the number of purchases of a buyer unchanged, but randomize the sellers (\textit{i.e.}, buyers buy from random sellers).
 We find that the increased communication between
 the two users does not correlate with the information passing success rate (\textit{Random} curve of Figure~\ref{fig:closure}(left)), meaning that the stronger the communication, the stronger the effect of information passing.

The results from this experiment lead to several observations.
First, messaging generally increases purchasing behavior in the Taobao network. Second,
information passing is present in the network. Additionally, communication
leads to purchases, and social network structure provides a surprisingly
strong signal indicating which seller a buyer will purchase from. This latter
result will be quite useful later when we study consumer choice prediction.

In addition to counting the number of messages exchanged between buyers, one
should also consider the time difference, $t_{2}-t_{1}$, between the initial trade and message from
\textit{$B_{1}$} to \textit{$B_{2}$}. We expect that the larger the time difference between the
initial purchase and message, the lower the influence of the message upon the
purchasing behavior of \textit{$B_{2}$}. As shown in Figure~\ref{fig:closure}(middle), the probability of
information passing success steadily decreases with time.
We observe the same effect regardless of whether we require the trade $(B_2, S_1)$ to be a
first time trade (\textit{FirstBuy Req}) or any trade
(\textit{Standard}). This implies that social product
recommendation and influence spreading is most effective when utilized
immediately after initial product adoption or purchase.

Although communication between \textit{$B_{1}$} and \textit{$B_{2}$} is a
significant influence upon information passing, the characteristics of the
product itself also affect the success of information passing. We can imagine
that the most important product attribute to consider is the price (in Chinese Yuan,
CNY) of the initial purchase.
As displayed in Figure~\ref{fig:closure}(right), we find that the information passing success rate decreases
with product price for the price range from 1 CNY to 15 CNY, then increases
slightly for products priced above 15 CNY. 
The large closure probability at a price of 1 CNY is
 due to the popularity and virality of virtual goods, such as game credits.

In addition to product price, we also consider the category of the product
itself when measuring the information passing success rate. We find that a few categories exhibit
recommendation success rates much higher than other categories, while many
categories exhibit nearly no information passing at all. For example,
the category \textit{women's clothing} exhibits a success
rate of over 20\%, while the category \textit{home decorations} exhibits a
success rate of 1.47\%. A major possible influence for the variability of these numbers are the regular group-purchasing
deals offered by large stores on Taobao, which provide financial incentives for consumers
to convince their friends to join them in purchases.

In summary, our experiments demonstrate that the success of information
passing is directly proportional to the communication strength
between the initial buyer and their message partner, inversely proportional to
the time difference between the purchase and initial recommendation, inversely
proportional to product price, and highly category specific.
It is clear that influencing an individual to purchase a product is a complex affair.
We believe these results are informative and can provide guidance to viral
marketing campaigns when trying to promote product recommendation and adoption among online users.

\begin{table*}[t]
  \centering
  {\small
  \begin{tabular}{c||r|r||r|r|r||r|r||c}
    \text{Dir. Config Set} & \text{\# Instances} & \text{\# Uniq. X} & \text{P(close)} & \text{P(t|close)} &
    \text{P(m|close)} & \text{s($t_{o}$)} & \text{s($t_{i}$)} & \text{X ``role''} \\
    \hline \hline
    \includegraphics[scale=0.10]{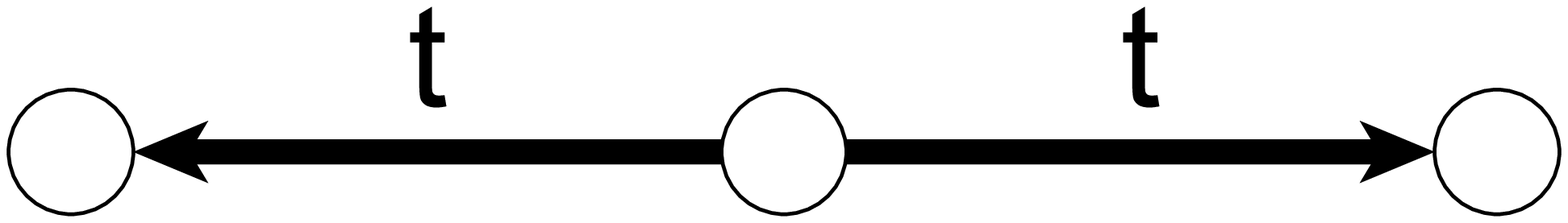} & 590,635 & 235,088 & 0.4146 & 0.4027 & 0.5973 & 69.19 & 63.39 & B\\
    \includegraphics[scale=0.10]{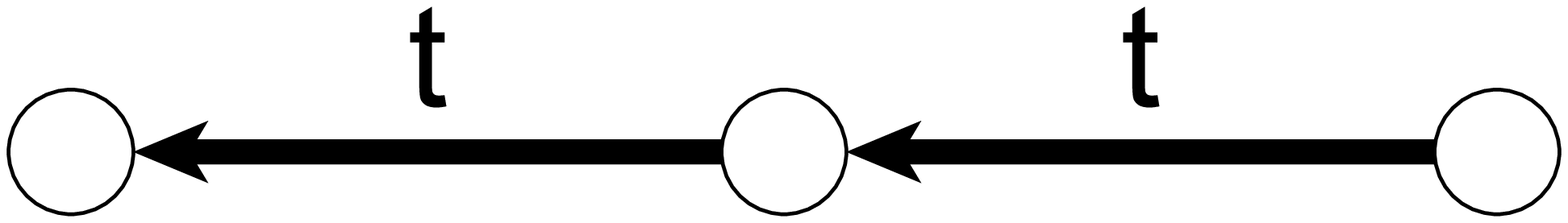} & 469,755 & 28,046 & 0.3925 & 0.3295 & 0.6705 & 3.09 & 16.87 & \\
    \includegraphics[scale=0.10]{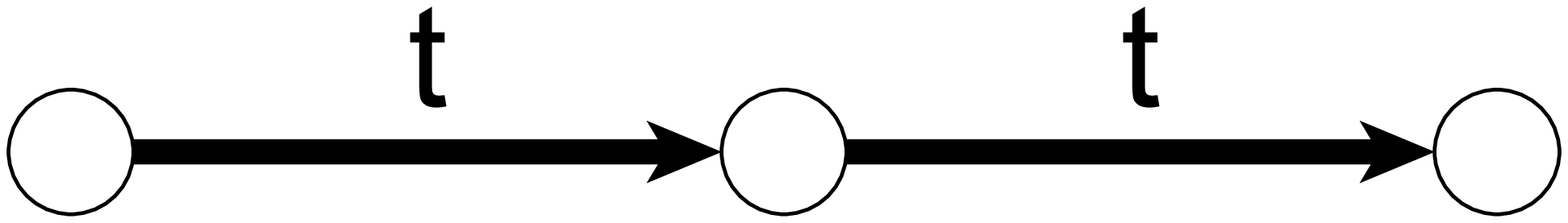} & 410,951 & 27,302 & 0.3319 & 0.3636 & 0.6364 & 18.75 & 6.13 & \\
    \includegraphics[scale=0.10]{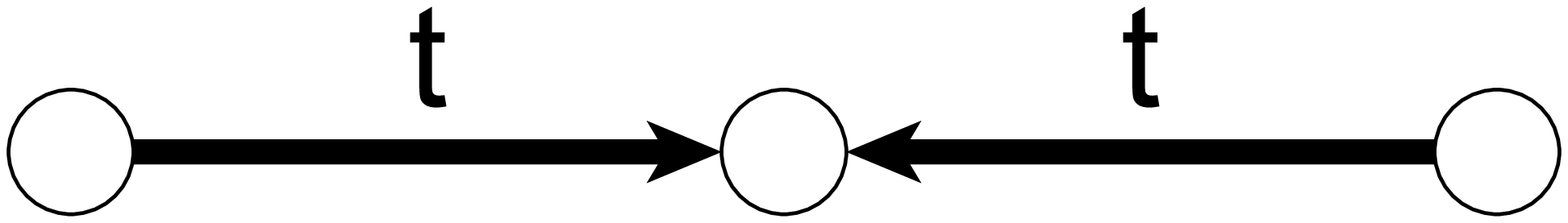} & 516,941,038 & 45,741 & 0.0018 & 0.1242 & 0.8758 & -18.11 & -18.30 & S\\
    \includegraphics[scale=0.10]{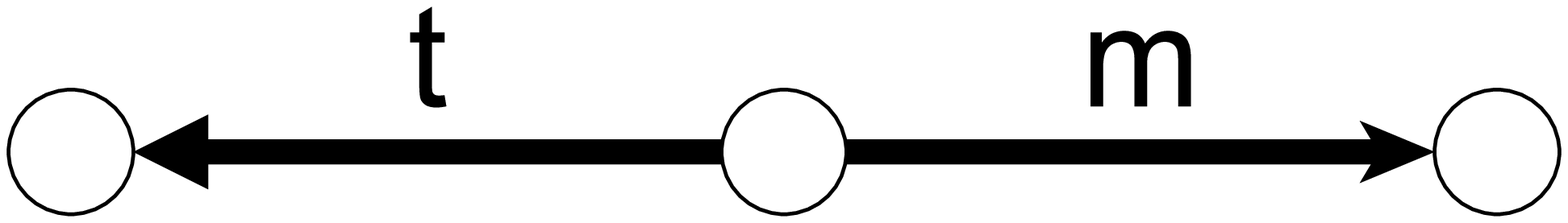} & 2,661,874 & 382,690 & 0.5034 & 0.3191 & 0.6809 & 41.26 & 101.61 & B\\
    \includegraphics[scale=0.10]{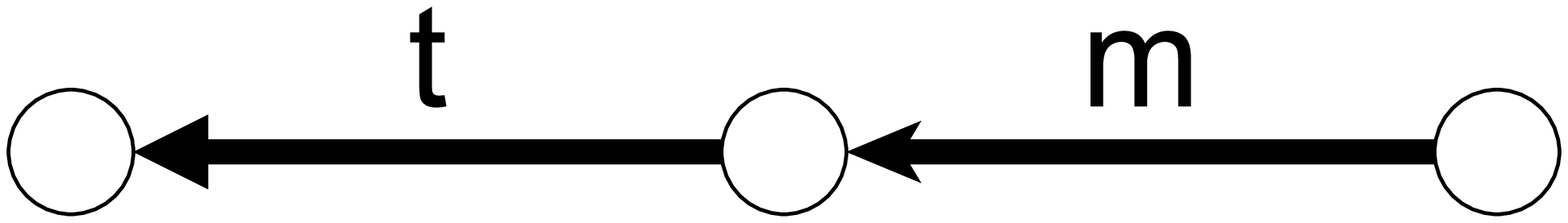} & 2,738,167 & 428,334 & 0.5470 & 0.3220 & 0.6780 & 41.94 & 118.83 & B\\
    \includegraphics[scale=0.10]{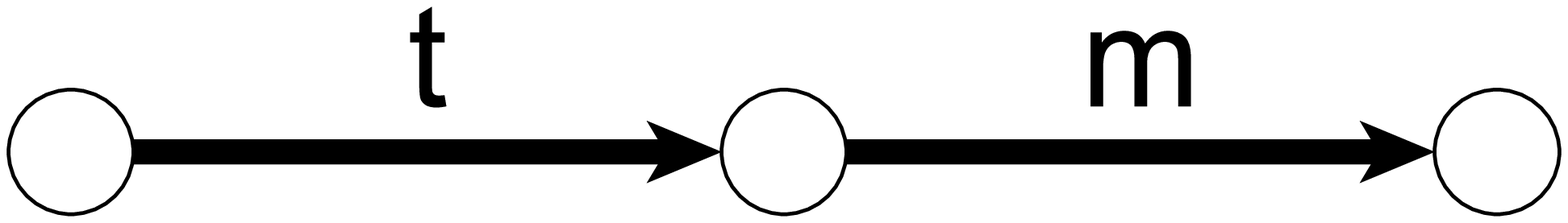} & 253,840,924 & 45,318 & 0.0048 & 0.1308 & 0.8692 & -9.28 & -5.40 & S\\
    \includegraphics[scale=0.10]{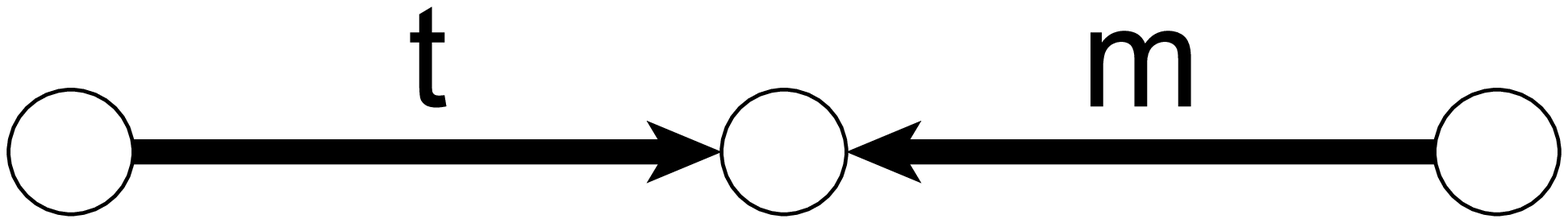} & 252,983,480 & 45,931 & 0.0050 & 0.1299 & 0.8701 & -9.26 & -5.03 & S\\
    \includegraphics[scale=0.10]{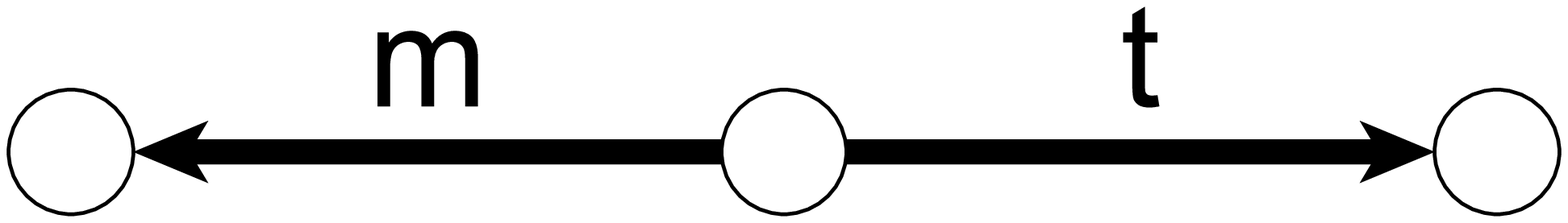} & 3,106,078 & 421,888 & 0.5103 & 0.3309 & 0.6691 & 126.17 & 61.89 & B\\
    \includegraphics[scale=0.10]{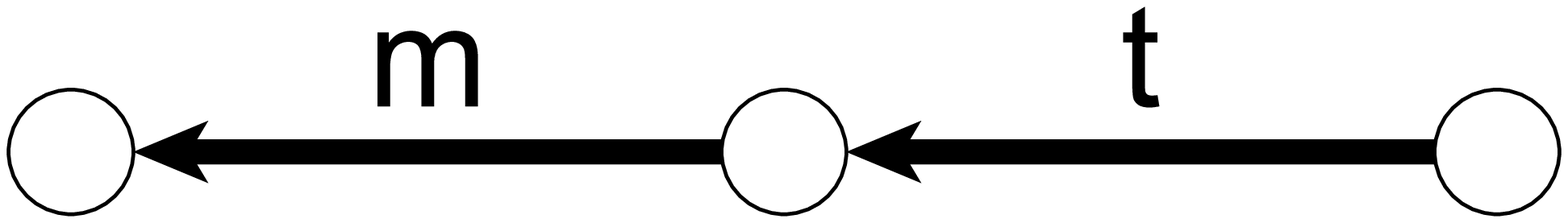} & 276,047,807 & 46,475 & 0.0070 & 0.1595 & 0.8405 & -0.59 & 2.09 & S\\
    \includegraphics[scale=0.10]{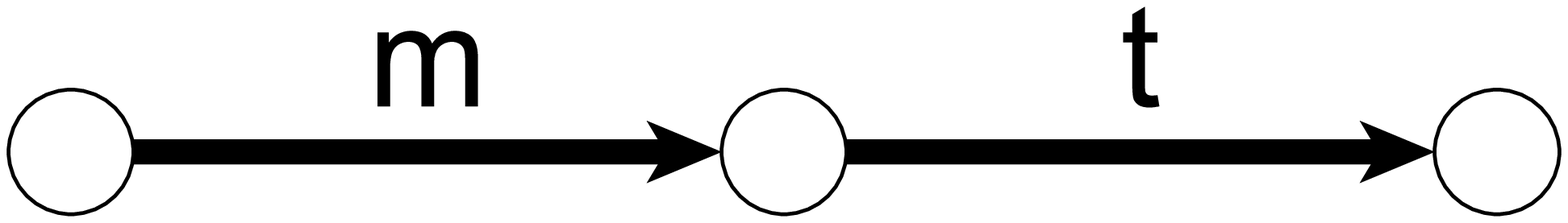} & 3,174,237 & 475,074 & 0.5019 & 0.3324 & 0.6676 & 141.02 & 65.23 & B\\
    \includegraphics[scale=0.10]{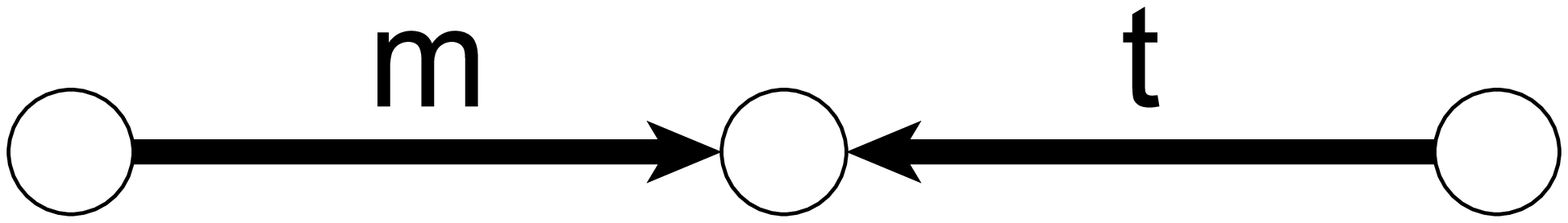} & 272,424,037 & 47,524 & 0.0070 & 0.1598 & 0.8402 & -0.04 & 2.63 & S\\
    \includegraphics[scale=0.10]{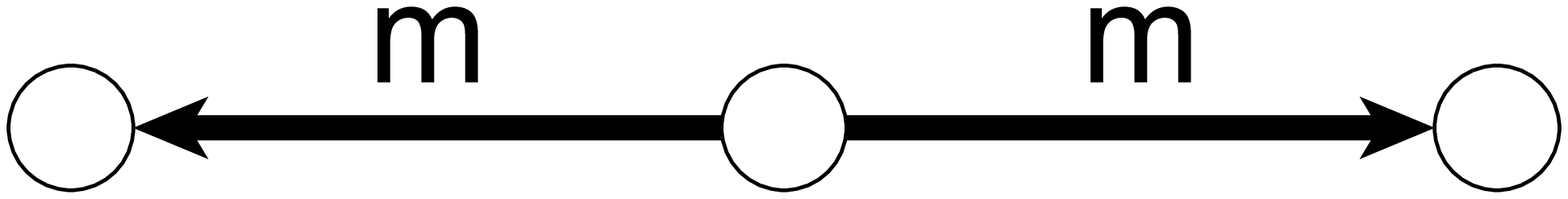} & 420,018,116 & 403,220 & 0.0289 & 0.1386 & 0.8614 & 52.92 & 63.96 & \\
    \includegraphics[scale=0.10]{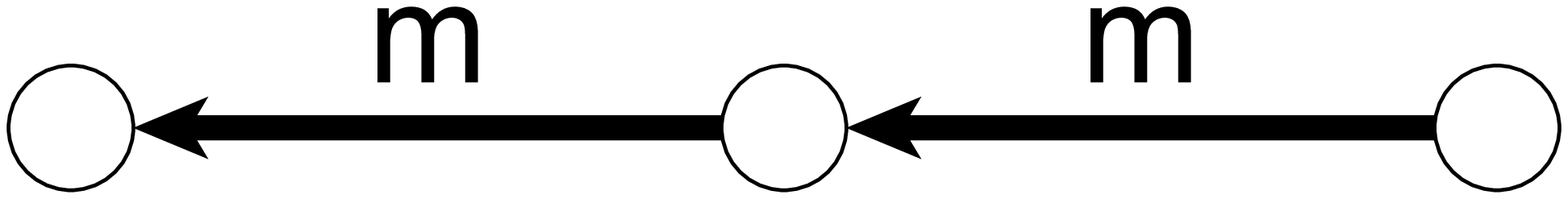} & 280,943,201 & 436,865 & 0.0458 & 0.1369 & 0.8631 & 56.03 & 75.16 & \\
    \includegraphics[scale=0.10]{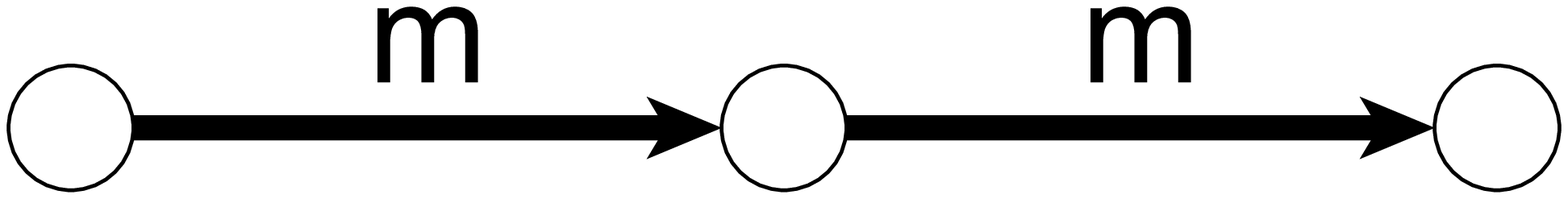} & 276,848,803 & 442,548 & 0.0438 & 0.1409 & 0.8591 & 67.61 & 70.19 & \\
    \includegraphics[scale=0.10]{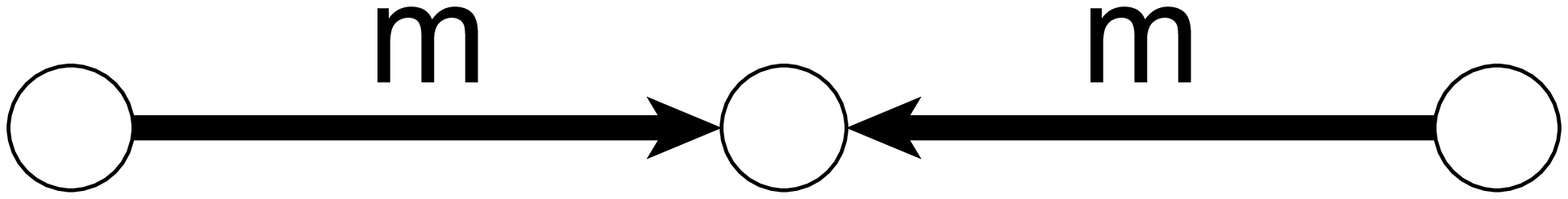} & 272,535,699 & 469,549 & 0.0467 & 0.1398 & 0.8602 & 68.69 & 80.00 &
    \end{tabular}
  }
    \vspace{-1mm}
  \caption{Directed Configuration Set (\textit{U}; \textit{X}; \textit{V}),
  where \textit{X} is the middle node, \textit{U} is the left node, \textit{V}
  is the right node. \text{\# Instances} = number of instances. \text{\# Uniq. X} = number of unique \textit{X} nodes over all
  instances. \text{P(close)} = 100 * probability of a triad-closing third edge. P(t|close)
  = proportion of triads closed by a trade. P(m|close) = proportion of triads closed by a message. s($t_{o}$) = surprise for directed trade edge
  (\textit{U}, \textit{V}). s($t_{i}$) = surprise for directed trade edge
  (\textit{V}, \textit{U}). \text{X ``role''} = hypothesized role of \textit{X}.}
  \label{table:dirTriads}
  \vspace{-3mm}
\end{table*}

\xhdr{Directed Triadic Closure}
Our previous study of local information passing processes can be seen as a
special case of the more general directed triadic closure problem.
We next investigate the process of directed triadic closure on a global scale.
In the following study, we answer various questions regarding triadic closure
including: What types of triads are more likely to be formed? What types of
edges are more likely to close triads?

For our task, we focus upon the link formation process in the
context of triadic closure for the message and trade networks. In particular,
consider a triple of nodes $U,X,V$, where first pairs $U,X$ and $V,X$ interact via
messaging or trading, and then a triangle closing edge $U\rightarrow V$ or
$V\rightarrow U$ forms. We are interested in how the type of interaction (message vs. trade) and the direction of interaction (i.e., $U\rightarrow X$
vs. $X\rightarrow U$) affects the formation of the triad-closing edge between $U$ and $V$.

Let us define a Directed Configuration Set (\textit{U}; \textit{X}; \textit{V})
as a situation where an edge forms between \textit{U} and \textit{X} at time \textit{$t_{1}$}, then an
edge forms between \textit{V} and \textit{X}
at time \textit{$t_{2}$} (\textit{$t_{2}$}$>$\textit{$t_{1}$}).
We are interested in the probability that a triad-closing edge forms between nodes
\textit{U} and \textit{V} at a time \textit{$t_{3}$} (\textit{$t_{3}$}$>$\textit{$t_{2}$}).
There are 16 possible Directed Configuration Sets, displayed in the
first column of Table~\ref{table:dirTriads}: the left node represents $U$, middle node $X$ and right node $V$.
For the triad-closing edge, we use the following shorthand notation: $m_{i}$ is the message edge (\textit{V},\textit{U}), while $m_{o}$ is the message edge in the opposite direction (\textit{U},\textit{V}). Similarly, $t_{i}$ is
the trade edge (\textit{V},\textit{U}), while $t_{o}$ denotes the trade edge in the opposite direction
(\textit{U},\textit{V}). We use the term \textit{instance} to refer to a particular example of a configuration set.

Now in Table~\ref{table:dirTriads}, we examine various properties of
configuration sets. In particular, we are interested in knowing, ``How many
times does a particular configuration set get closed with a third edge? And
what is the type of that edge?'' For easier reasoning about various
configuration sets, we denote hypothetical buyer/seller designations for the
middle node \textit{X} in the last column of Table~\ref{table:dirTriads}. For
example, if the configuration contains a purchase by \textit{X} and no sale, then we say that \textit{X} has a ``buyer'' role. Similarly, if the configuration contains a sale by
\textit{X} and no purchase, then we say that \textit{X} has a ``seller'' role. In all other cases, the role of \textit{X} is ambiguous.

To begin our study of directed triadic closure, first observe that there is little brokerage or reselling in the network, as the two configurations where \textit{X} both ``buys'' and ``sells''
have the lowest unique node $X$ as well as instance counts (Columns \textit{\# Uniq. X}, \textit{\# Instances}). This is indicative of
the general bipartite structure of Taobao, where users primarily take on either
buyer or seller roles.

Another important observation is that configurations where \textit{X} has a
``seller'' role are represented approximately 100
times more often than configurations where \textit{X} has a ``buyer'' role. The number of unique buyers exceeds the number
of unique sellers in these configurations, as shown in Column \textit{\# Uniq. X},
implying that activity levels for sellers are much higher than those for
buyers. The large difference in activity levels is likely due to how individuals actually use Taobao. Buyers browse Taobao casually and interact
with others primarily when interested, whereas sellers spend their day speaking
with potential clients. Given the bipartiteness of Taobao and the general
activity level of sellers, we can imagine that seller nodes are local ``star'' structures in the Taobao graph.

Following our comparison of configuration instance counts, we consider the
question, ``When will an instance of a Directed Configuration Set be closed by
a third edge?'' We compute the probability of the configuration (\textit{U}; \textit{X};
\textit{V}) being closed by an edge (\textit{U},\textit{V}).
Observe that configurations where \textit{X} is
a buyer have much higher closure probabilities (average 0.0051) than configurations where
\textit{X} is a seller (average 0.000046).
The large difference in closure probabilities is due to the fact that triads with middle buyers primarily consist
of two buyers and one seller, with the required third edge being a
buyer-seller edge. In contrast, triads with middle sellers likely contain one
seller and two buyers, with the required third edge being a buyer-buyer
edge. Since the Taobao network is essentially a bipartite network of buyers and
sellers, buyer-seller edges occur much more often than buyer-buyer
edges, leading to triadic closure for buyers being over 100 times more likely than for sellers.

\xhdr{Triad-Closing Edge Type Distribution}
After computing the instance counts and triad-closing
probabilities of each Directed Configuration Set, we next examine the distribution of edge types closing each type of configuration. For
each of the 16 configurations $c_{i}$, we count the number of instances that are closed by messages and trades in each
direction. Column \textit{P(m|close)} of Table~\ref{table:dirTriads} shows that
messages close most of the triads in the network. However, messages are also approximately 3 times as common as
trade edges in the data. Therefore, we need to compute expectations for each of the 4 possible triad-closing edge types ($m_{i}$, $m_{o}$, $t_{i}$, and $t_{o}$).

We define a node's \textit{generative baseline} as the proportion of its
out-edges that are trades. We assume that when a node $A$ creates an
edge, it generates a trade edge with a probability equal to its generative
baseline, denoted by $p_{t}(\textit{A})$.
For the configuration $c_{i}$, the expected number of triads that
are closed by a trade from $U$ to $V$ is equal to $\sum_{U \in c_{i}} p_{t}(U)$,
where the summation is over all instances of the configuration $c_{i}$. Similarly, the expected number of triads that are closed by a trade from $V$ to $U$
is $\sum_{V \in c_{i}} p_{t}(V)$. Viewing each instance of edge generation as a
separate Bernoulli trial, we derive an expression for \textit{surprise}~\cite{leskovec2010signed} to indicate the number of signed standard deviations
by which an observed edge type count differs from expected.
For the configuration $c_{i}$, the surprise of the triad-closing edge (\textit{U},\textit{V}) being
a trade is $s_{t_{o}}$ = $\frac{\#observed - \sum_{U \in
c_{i}}p_{t}(U)}{\sqrt{\sum_{U \in c_{i}} [p_{t}(U)*(1-p_{t}(U))]}}$ , where
\#observed is the number of instances of $c_{i}$ that are closed by a trade
edge (\textit{U},\textit{V}). Similarly, we compute the surprise $s_{t_{i}}$ of the triad-closing edge
(\textit{V},\textit{U}) being a trade. These trade surprise values are listed
in Columns \textit{s($t_{o}$)}, \textit{s($t_{i}$)} of Table~\ref{table:dirTriads}.\footnote{We do not explicitly compute the surprises for directed message edges
since $p_{m}(X) = 1 - p_{t}(X)$, implying that s($m_{o}$) = -s($t_{o}$) and s($m_{i}$) = -s($t_{i}$).}

After computing these edge type surprises, we can compare observed
triad-closing edge counts with expected edge counts from our generative
baseline. Our first observation regarding directed edge surprises is that for configurations where $X$ is a buyer, the triad-closing edge being a trade edge is observed significantly more than
expected, as shown in Table~\ref{table:dirTriads}. An explanation for this is that configurations with middle buyers primarily consist
of two buyers and one seller, with the required third edge being a
buyer-seller edge. Since Taobao is essentially a bipartite network, trade surprises for such
configurations vary from 40 to 140 standard deviations more than
expected.

For configurations where $X$ is a buyer and only one of
$U$ and $V$ is a seller, s($t_{o}$) and s($t_{i}$)
differ by a factor of 2. In particular, the trade surprise of the edge directed
toward the seller is twice as large as the other direction. This is an
example of how the role of each of the nodes in a configuration influence both the
edge type and edge direction probabilities of the triad-closing third edge.

In contrast, for configurations where $X$ is a seller, both trade surprises,
s($t_{o}$) and s($t_{i}$), are negative.
As previously mentioned, when $X$ is a seller, $U$
and $V$ are likely to be both buyers. Since buyers rarely purchase from
each other on Taobao, this leads to messages being observed more than
expected between $U$ and $V$.

Our analysis of edge type surprises indicates the significance of user roles in
dictating edge formation in the Taobao network. Message edges close the
majority of the triads in Taobao due to their relative proportion among all
network edges. However, the relative proportion of triad-closing edges being a
message edge is primarily dictated by the user role of the middle node $X$.

\section{Price of trust}
\label{sec:trust}
\begin{figure}[t]
\centering
\includegraphics[width=0.40\textwidth]{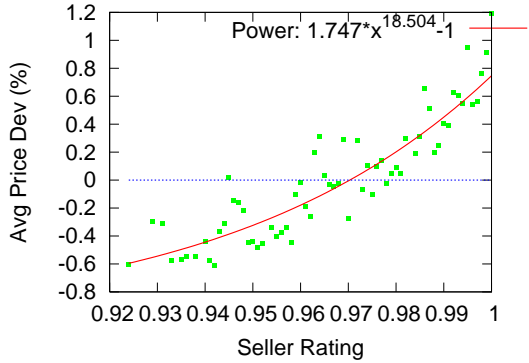}
  \vspace{-3mm}
\caption{Per Item: Average price deviation from median (\%) vs seller rating(\%).}
\label{fig:item_vs_price}
    \vspace{-3mm}
\end{figure}

\begin{figure}[t]
\centering
\includegraphics[width=0.40\textwidth]{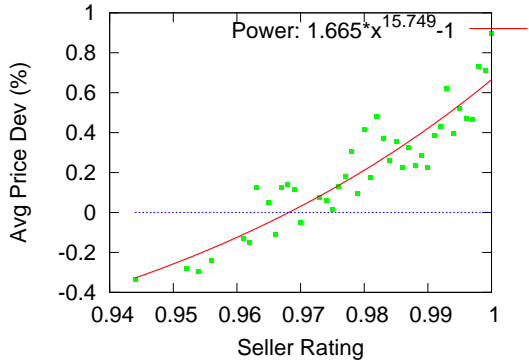}
  \vspace{-3mm}
\caption{Aggregated Per Seller: Average price deviation from median (\%) vs seller rating(\%).}
\label{fig:seller_vs_price}
    \vspace{-3mm}
\end{figure}

The prior study of information passing relies upon the spread of influence
through buyer-buyer communication, which can be seen as an implicit form of
buyer-buyer trust. We next examine a more explicit form of buyer-seller trust
encapsulated by seller ratings. In the context of electronic marketplaces, buyers are unsure about seller trustworthiness,
so buyers put their trust into seller ratings and reviews, and are willing to
pay a premium to sellers with good reputations~\cite{resnick2006value}. 
How much extra will a buyer pay for transaction with a highly rated seller?

\xhdr{Data Preparation}
To answer this question, we use the large Taobao transaction dataset to
study how good seller reputations are rewarded on Taobao and quantify a price
for trust. To facilitate our experiments, we perform a web crawl of the Taobao
website to obtain product and seller metadata associated with the transactions in our
original dataset.
Each transaction in Taobao is rated by the buyer, so
we use the percentage of positive reviews that each seller has received in
the past as a proxy for seller reputation and trustworthiness. Henceforth, we
shall refer to that ratio as seller \textit{rating}.

With this rating information, we compare sellers of the same product
and determine how their sale prices differ. The difficult step of the
experimental setup is identification of all product listings in our dataset
which correspond to the same products.
We develop a high precision method
targeted toward specific types of products and use our method to identify and
group together product listings referring to the same product into \textit{product clusters}.
The resulting dataset for our study of trust consists of 382,980
items, 11,293 product clusters (corresponding to unique product types), and 6,199 unique sellers.
Each of these product clusters contains a set of items which correspond to the
same exact product. 

\xhdr{Quantifying Trust}
With our product cluster dataset, we study the relationship between seller rating and the price at which a seller can transact.
For each product cluster, for each listing within a cluster, we compute \%
deviation from median cluster price. We plot the average price deviation from
the median per listing versus seller rating in Figure~\ref{fig:item_vs_price},
and fit the data with a power function ($R^{2}$ of 0.80). From the super-linear fit, we see that a higher
rating is associated with a seller selling his products at a premium compared to most of
his peers. 
Another interesting observation is that a seller rating of 97.1\% corresponds to transaction at the median cluster
price. The lack of negative reviews in such data has been observed in other
e-commerce data such as eBay~\cite{resnick2002trust}, and has been hypothesized
to be the result of a ``high-courtesy'' social norm.

Next, we aggregate the average price difference of all items sold by a seller and plot that against
seller rating, shown in Figure~\ref{fig:seller_vs_price}. The primary difference between this plot and the previous one is
that we now aggregate all the different instances of a seller's transactions within the
same cluster, and compare the sellers in a cluster against each
other.\footnote{There is large variation between activity levels among
  different sellers, so we only display those who have sold at least 15 items.} We find that a power function fits our
data particularly well ($R^{2}$ of 0.87). Looking across all sellers, the
elasticity of product price with respect to seller rating is a small, positive quantity, indicating that there is a direct relationship between seller
rating and increased sales price. 

One possible explanation for our findings is that highly rated
  sellers incur higher costs associated with their products, hence they can
  sell their products at a price premium. For example, highly rated sellers
  may provide better services, such as replying to messages from customers
  in a timely fashion, or shipping products more frequently. An alternative explanation is that buyers are willing to pay
  more to highly rated sellers to minimize transaction risk, thus sellers who
  maintain good reputations are financially rewarded. Although higher seller ratings
  are correlated with higher sales prices, the small magnitude of the
  elasticity indicates that buyer purchasing decisions are likely influenced by
  other variables, such as the social network in which the purchases are embedded.
This leads us to consider the scenario of consumer choice prediction.

\section{Consumer choice prediction}
\label{sec:predict}
\begin{figure}[t]
\centering
\includegraphics[width=0.23\textwidth]{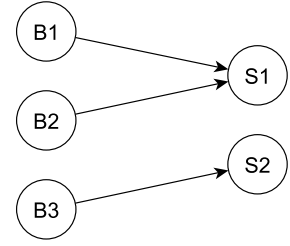}
  \vspace{-3mm}
\caption{Buyer-Seller Cluster. Given that S1 and S2 both sell exactly the same product that
  the buyers buy, predict the correct seller for each buyer B1, B2, B3.}
  \vspace{-5mm}
  \label{fig:bipartite}
\end{figure}

In order to demonstrate the power of network structure, we now consider the problem of {\em consumer choice prediction}.
Imagine the following situation: a user comes to Taobao and issues an exact
query for the product (s)he aims to buy. There is a list of $k$ sellers selling
the exact product the buyer is going to buy. Which seller will the buyer purchase
the item from? The seller with the lowest price? The most trusted seller?
The seller who interacted with the buyer's friends in the past?

Given that our prior experiments have demonstrated the importance of the social network and the strong presence of
information passing, we aim to investigate the role that social and communication networks play in consumer
decision making~\cite{degeratu2000consumer}. If the social network has no influence upon buyer behavior, then we expect that buyers
will purchase from sellers that offer the lowest price. However, as we will see, it is exactly the
social network information that gives the strongest signal in predicting which
seller a buyer will make a purchase from.

For this consumer choice prediction task, we use the product
cluster data described previously in our study of trust. Each product cluster
consists of a set of transactions between different pairs of people, but
corresponding to the exact same product.
We convert each product cluster into a bipartite subgraph composed of buyer and seller nodes, as
shown in Figure~\ref{fig:bipartite}, where the buyers and sellers have all
either bought or sold the same particular product. We will term these bipartite subgraphs
\textit{buyer-seller clusters}, and henceforth shall perform prediction with these
clusters.

Typically a seller will transact with multiple buyers
in the same cluster, while a buyer purchases the product from a single seller in the cluster. Both buyers and sellers can be
included in more than one buyer-seller cluster if they buy or sell multiple types of products. We restrict our focus to buyer-seller
clusters with at least 2 and no more than 10 sellers. Overall, our prediction
data is composed of 9,950 clusters, with a per-cluster average of 5.91 buyers
and 3.l3 sellers. It is important to understand that, by construction, all sellers
offer a product for sale that is exactly relevant to the buyer's interests. The task
now is, within a cluster, given a particular buyer, rank the sellers such that the seller the
buyer is going to buy from is ranked as high as possible.

\textbf{Problem Statement}: For each buyer-seller cluster \textit{$C_{i}$},
there is an associated set \textit{$B_{i}$} of buyers and set \textit{$S_{i}$} of
sellers. For each cluster \textit{$C_{i}$}, for each buyer \textit{$B_{ij}$} in
\textit{$B_{i}$}, predict which seller(s) from \textit{$S_{i}$} the buyer
\textit{$B_{ij}$} will purchase their product(s) from.

We model this prediction problem as a ranking problem, where for each buyer
\textit{$B_{ij}$} in cluster \textit{$C_{i}$}, we wish to generate a ranking of
the sellers \textit{$S_{ik}$}, such that the true seller from whom
\textit{$B_{ij}$} actually purchased the product has the highest rank (i.e., score).
Since the positive and negative examples for our problem
come in sets, it is natural to use a ranking based machine learning approach.
In particular, we use the Support Vector Machine SVM-rank~\cite{joachims2006training}.

\begin{table*}[t]
 \centering
{ \small
\begin{tabular}{l|l|l||l|l|l}
Feature Type & Feature Name & Feature Description & T & M & C\\ \hline \hline
\multirow{5}{*}{Product Metadata Features} & Fractional Price Rank & Seller
ranking using their median product price & & & \\
 & Fractional Rating Rank & Seller ranking using their rating percentage & & & \\
 & Historical Sold & Num. of all products sold by seller. 1) fractional rank
2) log of value & & & \\
 & Inventory Sold & Quantity already sold in the particular product listing. & &
& \\
 & Insurance & If the product is insured by the seller & & & \\ \hline
\multirow{4}{*}{Direct Network Features} & Buyer-Seller Interactions & 1) Trade
volume 2) Message volume 3) Are they contacts? & X & X & X\\
 & Time Since Last Transaction & Computed for both message and trade networks
&X & X &\\
& Fractional Message Rank & Seller ranking using number of buyer-seller
messages & & X & \\
& Nodal Trade Volumes & Number of trades for buyer and seller in the 2 month observation
period & X & & \\ \hline
\multirow{4}{*}{Indirect Network Features} & Number of mutual partners & Number
of nodes who have messaged or transacted with both buyer and seller & & X& X\\
 & Seller Clustering Coefficients & Computed for both message and contact
networks & & X & X\\
 & Mutual Densities & Frac. of edges between the set of nodes
mutual with both buyer and seller & & X & X\\
 & Seller PageRanks & Computed for static endtime networks & X & X & X
\end{tabular}
}
  \vspace{-3mm}
  \caption{Feature Set. T, M, C denote if the feature is computed on the trade, message,
  and contact networks, respectively.}
  \label{table:featureset}
  \vspace{-3mm}
\end{table*}

For consumer choice prediction, we use a base set of 23 features that describe product, buyer and seller metadata. We also use features that describe the buyer-seller interactions and network structure.
Table~\ref{table:featureset} lists the features we use in our experiments,
along with the networks they are computed on.\footnote{For each
  buyer decision, network features are computed from the snapshot of the
  network which existed the day prior to the true purchase date. This is
  necessary to properly model buyer decisions.}

\xhdr{Experimental Setup and Evaluation}
Our data consists of 58,812 sets of training examples (i..e, buyer-seller pairs where a buyer can
buy the same product from multiple sellers). We split the data into 75\% train and 25\% test sets.
The SVMs were trained with linear kernels and loss functions were chosen to minimize the number of incorrect
constraints. Since we typically have one positive example per buyer decision, this is
equivalent to optimizing Precision@1.

As a point of comparison for our models, we construct three simple
rule-based baselines:\\
\indent $\bullet$~\textbf{Random} baseline - rank the sellers randomly \\
\indent $\bullet$~\textbf{MinPrice} baseline  - rank the sellers by increasing price.\\
\indent $\bullet$~\textbf{MostMsg} baseline - rank the sellers by decreasing
buyer-seller message volume. Defaults to Random if no message edge is
present.

We evaluate the models and baselines using the following three
metrics:\footnote{Performance was also evaluated with several other standard ranking metrics. Results are similar, so hence not displayed.}\\
\indent $\bullet$ \textbf{Precision at Top 1} (P@1) - Fraction of times that
the top ranked seller is actually the true seller. (Higher is better) \\
\indent $\bullet$ \textbf{Mean Rank} (MR)- Average \textit{Rank} of the true seller. (Lower is better) \\
\indent $\bullet$ \textbf{Mean Reciprocal Rank} (MRR) - Average
\textit{Reciprocal Rank} ($\frac{1}{\textit{Rank}}$) of the true seller. (Higher is better)

\begin{figure*}[t]
  \centering
  \subfigure[P@1: SVMs vs Baselines]{\includegraphics[width=0.24\textwidth]{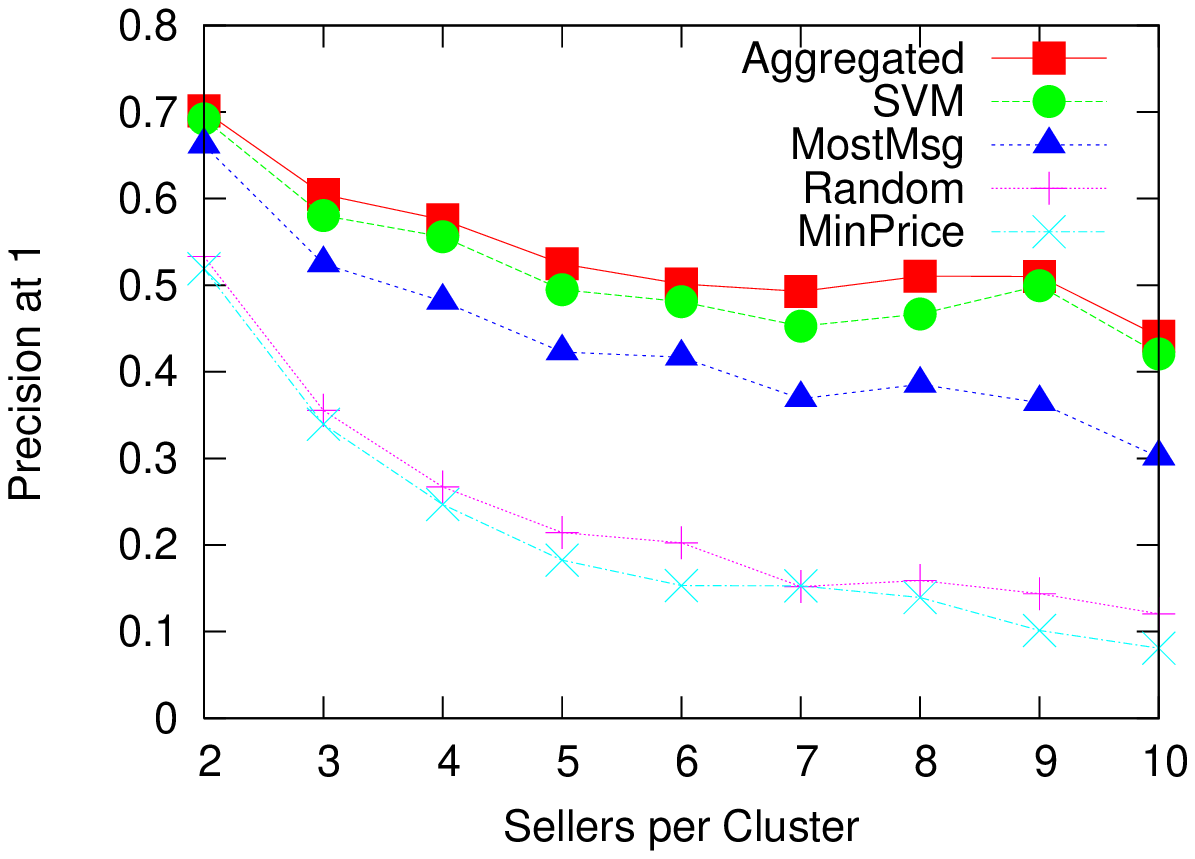}}
  \subfigure[P@1: Edge Types]{\includegraphics[width=0.24\textwidth]{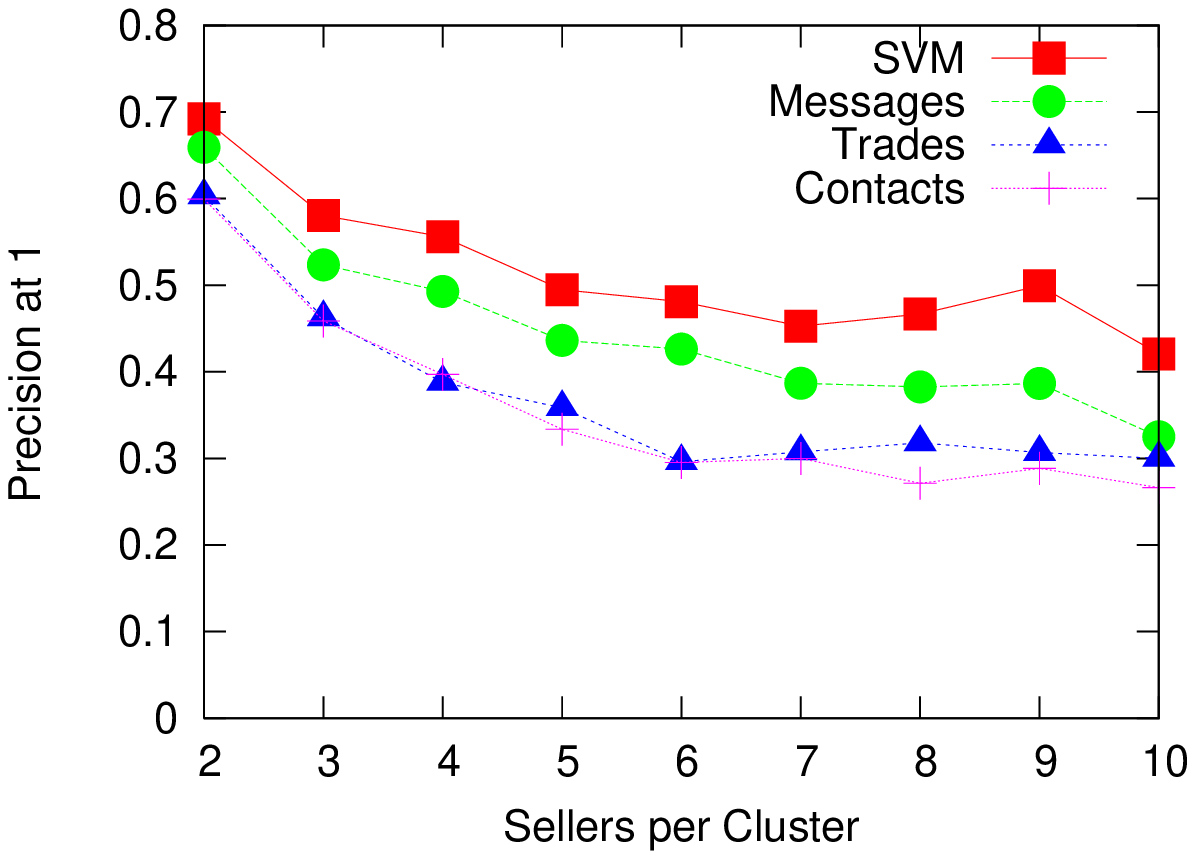}}
  \subfigure[P@1: Direct vs Indirect]{\includegraphics[width=0.24\textwidth]{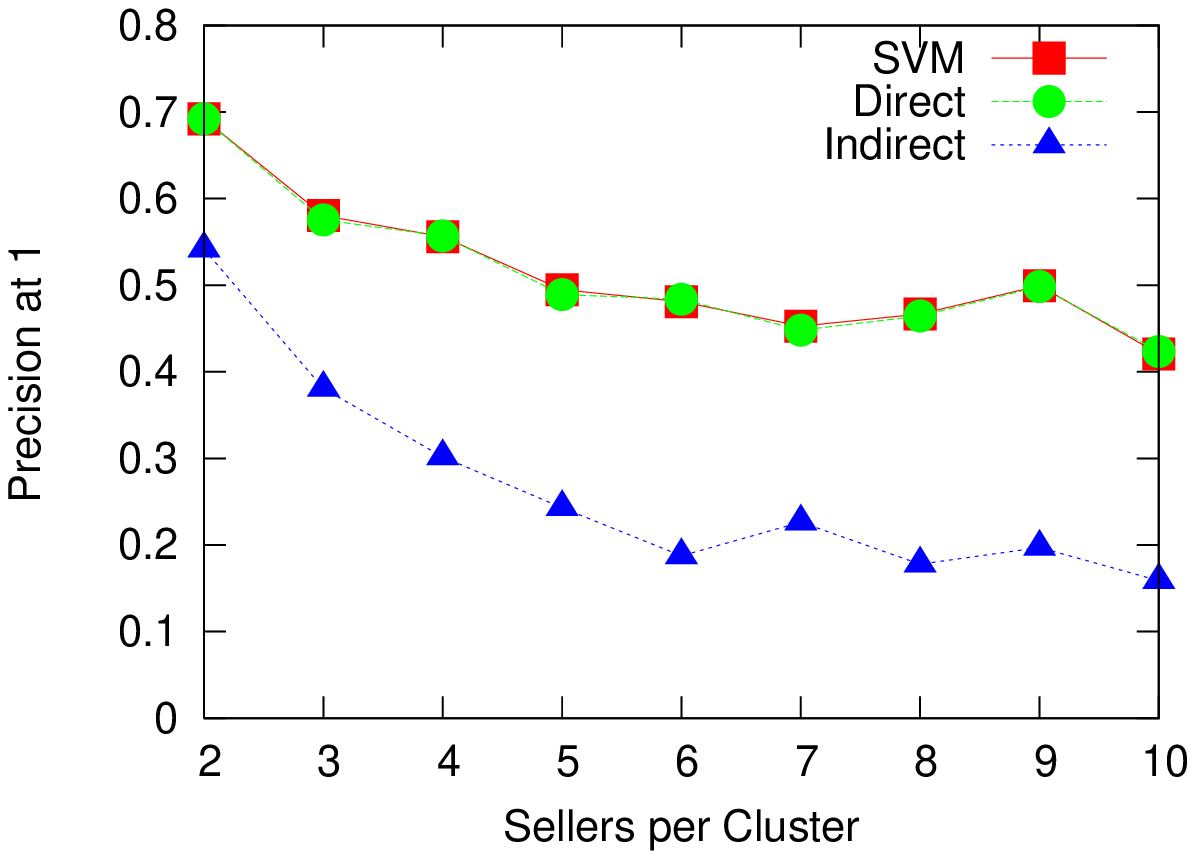}}
  \subfigure[P@1: Graph vs Metadata]{\includegraphics[width=0.24\textwidth]{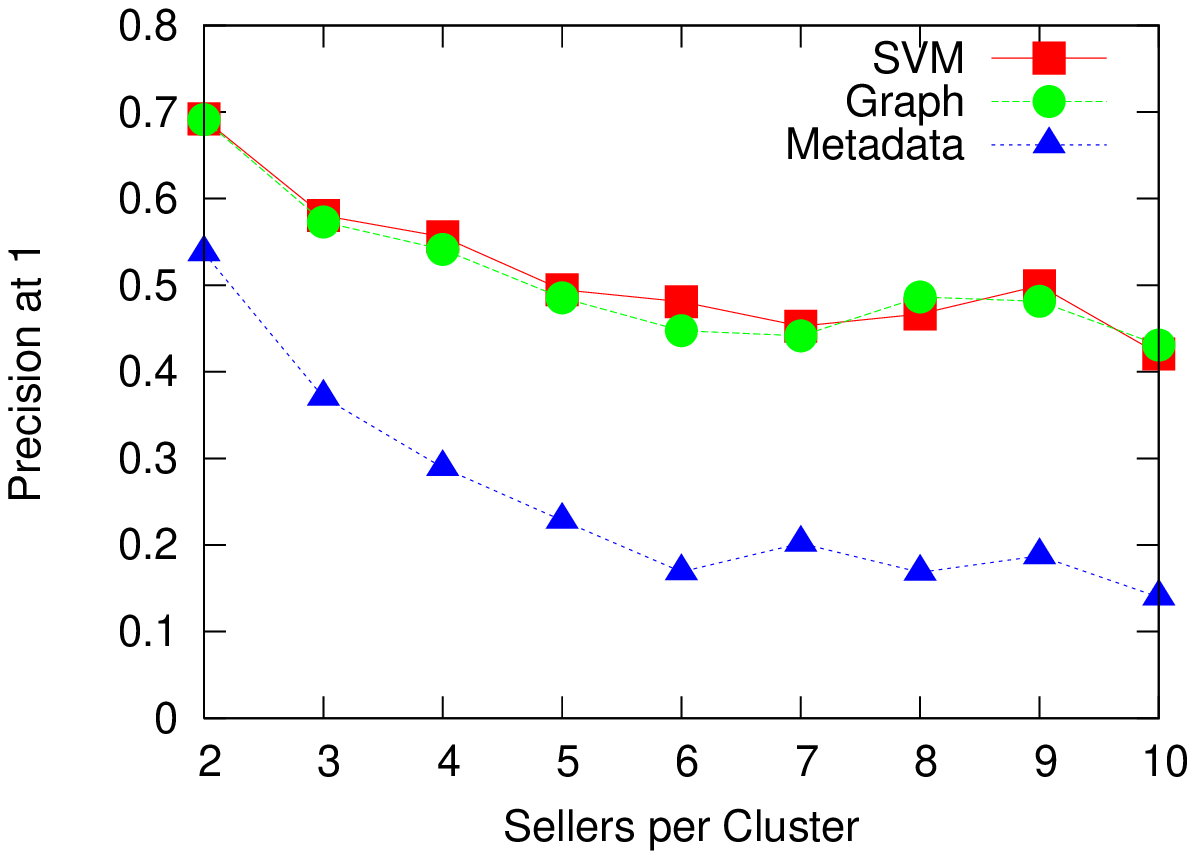}}
  \subfigure[MRR: SVMs vs Baselines]{\includegraphics[width=0.24\textwidth]{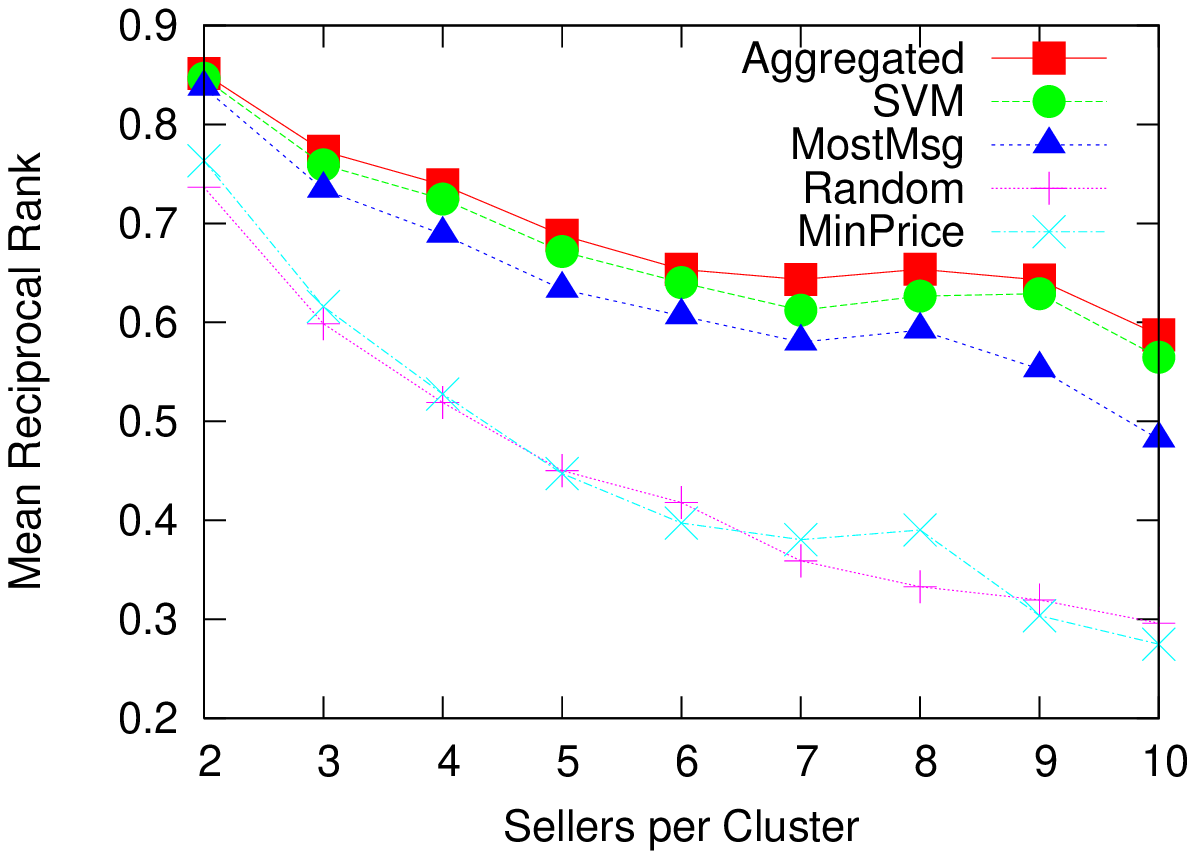}}
  \subfigure[MRR: Edge Types]{\includegraphics[width=0.24\textwidth]{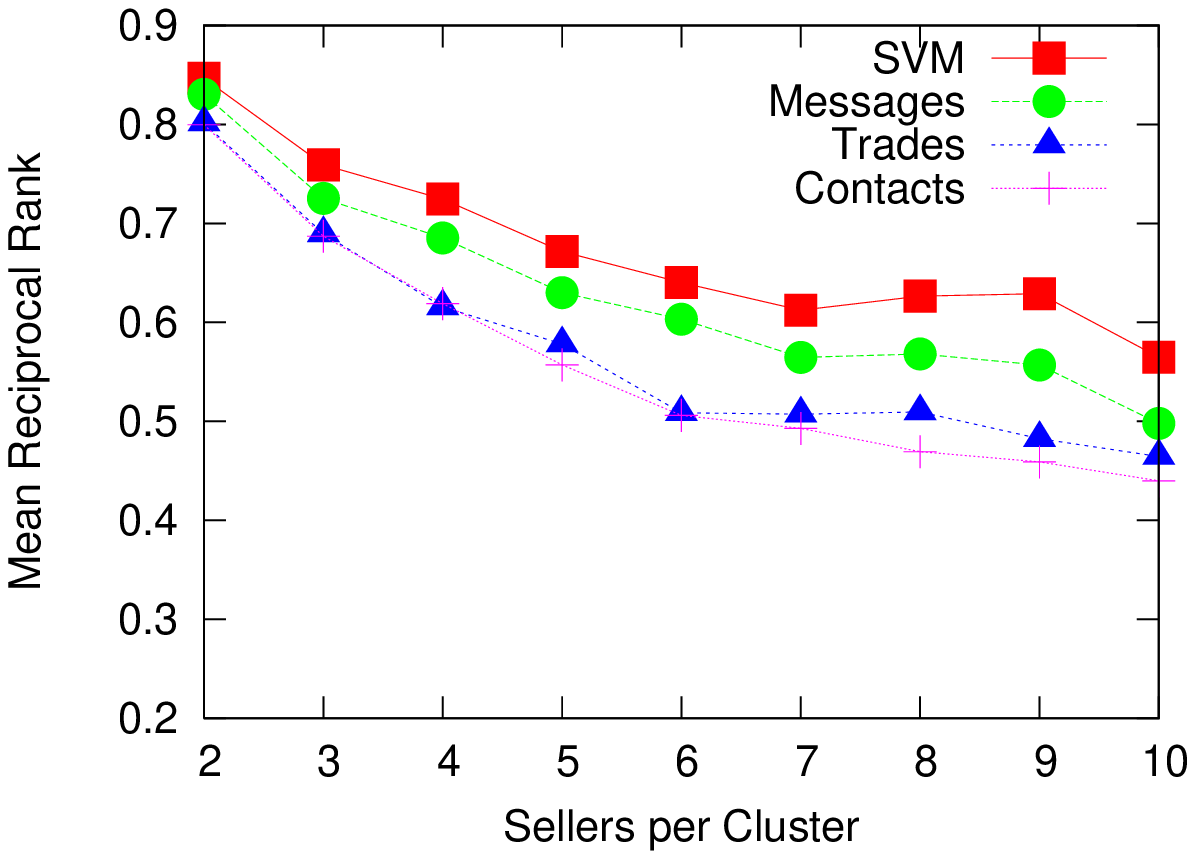}}
  \subfigure[MRR: Direct vs Indirect]{\includegraphics[width=0.24\textwidth]{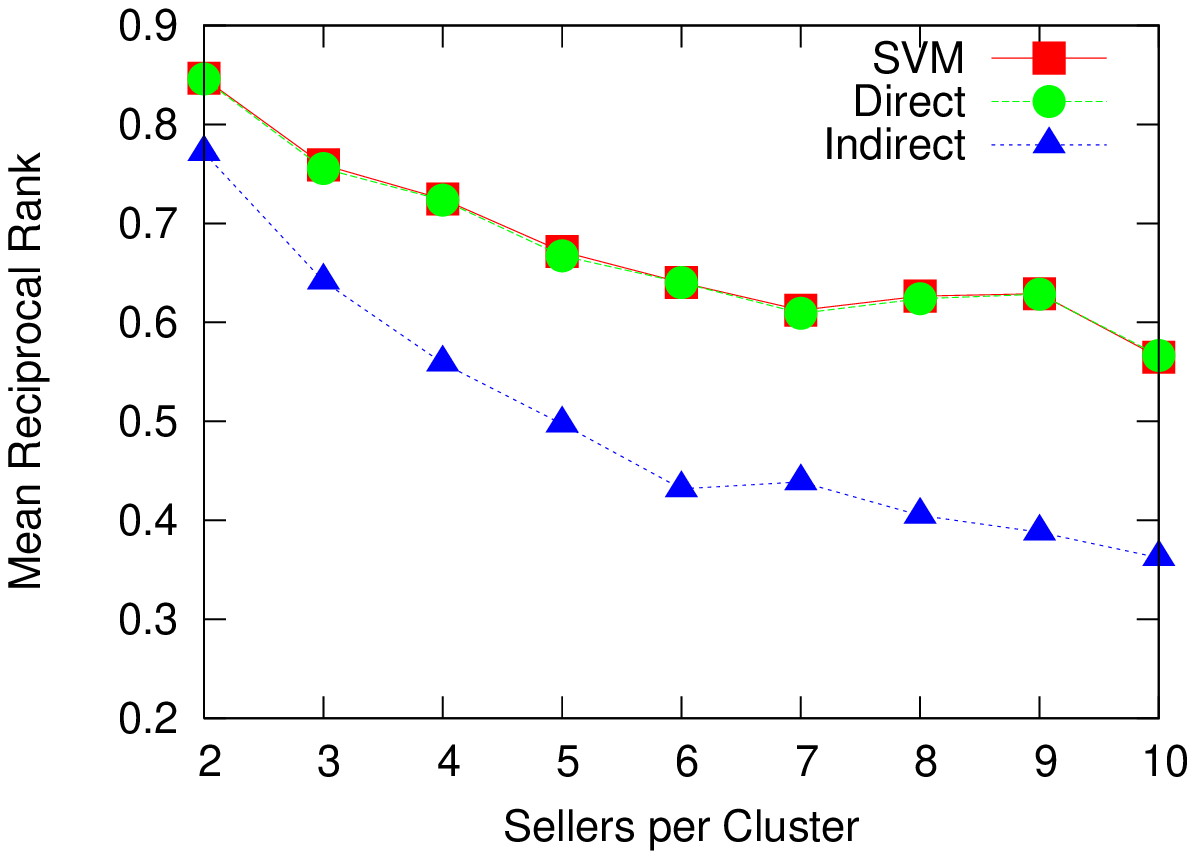}}
  \subfigure[MRR: Graph vs Metadata]{\includegraphics[width=0.24\textwidth]{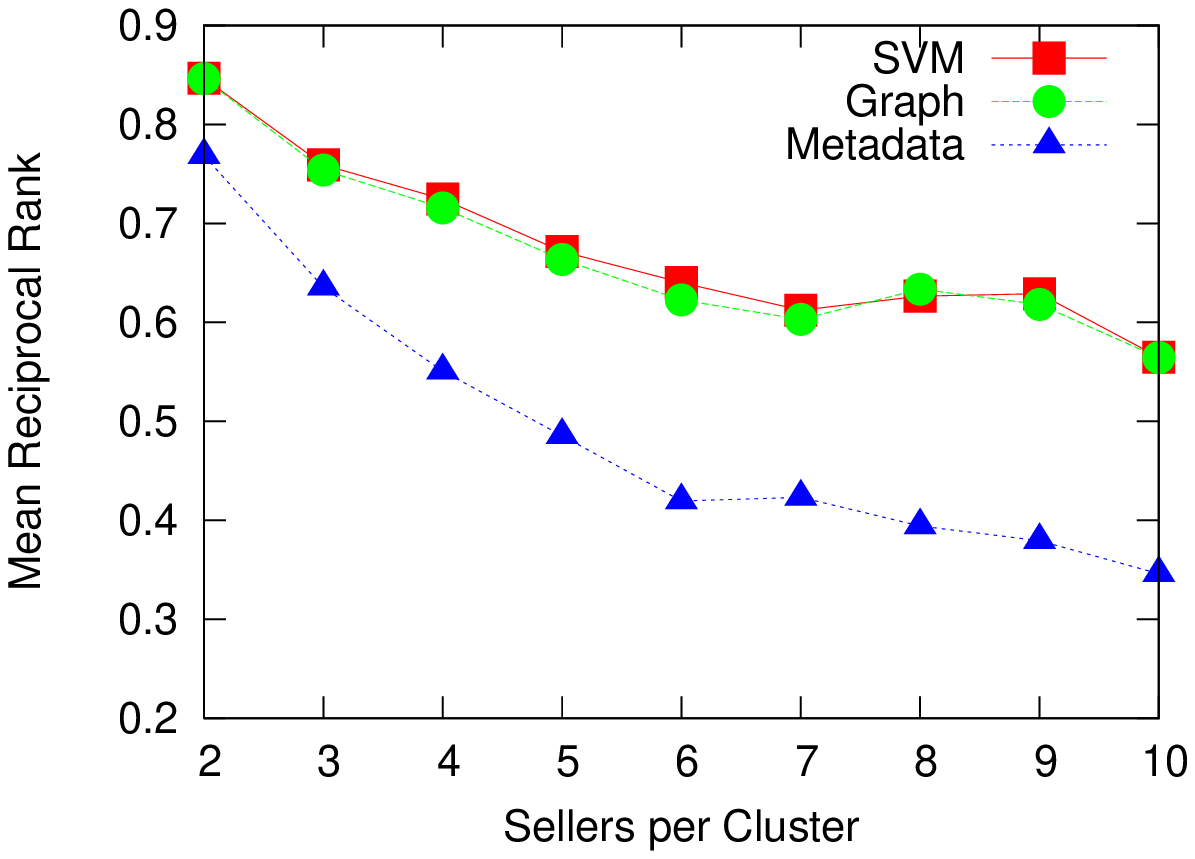}}
    \vspace{-3mm}
  \caption{Consumer choice prediction performance. The social graph is the most
  important feature in predicting which seller a buyer will purchase from.}
  \label{fig:prediction}
  \vspace{-5mm}
\end{figure*}

\begin{table}[t]
  \centering
  \begin{tabular}{r||c|c|c}
    \text{Feature Set} & \text{P@1} & \text{MRR} & \text{MR}\\
    \hline \hline
    \text{All Features} & 0.56 & 0.72 & 2.00\\
    \hline
    \text{Only Network} & 0.55 & 0.72 & 2.02\\
    \text{Only Meta} & 0.33 & 0.57 & 2.62\\
    \hline
    \text{Meta + Msgs} & 0.50 & 0.69 & 2.13\\
    \text{Meta + Trades} & 0.43 & 0.64 & 2.34\\
    \text{Meta + Contacts} & 0.42 & 0.63 & 2.40\\
    \hline
    \text{Meta + Direct} & 0.56 & 0.72 & 2.00\\
    \text{Meta + Indirect} & 0.34 & 0.58 & 2.58\\
    \hline \hline
    \text{MostMsg} & 0.50 & 0.69 & 2.11\\
    \text{Random} & 0.31 & 0.53 & 2.90\\
    \text{MinPrice} & 0.29 & 0.54 & 2.77
 \end{tabular}
   \vspace{-3mm}
    \caption{Customer choice prediction results.}
  \label{table:predResults}
    \vspace{-5mm}
\end{table}

\xhdr{Experimental Results}
Table~\ref{table:predResults} gives an overview of our experimental results where we compare the models using various feature sets and the baselines.
First, we note that the model trained on all 23 features gives a 79\% improvement over the
P@1 of the Random baseline and a 38\% improvement over Random's
MRR. The model also gives a 13\% improvement over the
MostMsg baseline and a 93\% improvement over the
MinPrice baseline. For all our evaluation metrics,
the model displays significantly better performance than the 3 baselines.
It is interesting to note the poor performance of MinPrice compared to
MostMsg. This suggests that communication links and the social graph are essential to understanding how
consumers make purchasing decisions in social commerce networks.
Note that a natural explanation for the importance of the social graph can be that a buyer first messages a seller, then immediately trades with him. However, we control for this by discarding
all communication on and after the trade date.

To evaluate prediction performance in more detail, we graph the P@1 and MRR of
the Full model (labeled as \textit{SVM}) and the 3 baselines versus the number of
sellers in the buyer-seller cluster (i.e., how many different sellers a buyer can choose
from) in Figures~\ref{fig:prediction}(a),(e). As expected, the performance of
all models and baselines decreases as the number of sellers in the cluster
increases. Observe that the performance gap between the Full model and MostMsg, the strongest baseline, widens as
the number of sellers to choose from increases and prediction becomes more difficult. If we look at the proportional P@1 improvement of the Full model over
MostMsg, we see only a 4.5\% improvement when there are 2 sellers, but a
39.5\% improvement when there are 10 sellers. In particular, we would like to highlight the
Full model's strong P@1 of 42.1\% for the challenging prediction task with 10
sellers. 
In general, the full power of the model is not realized until the prediction
problem becomes difficult for simple rule-based heuristics.

\xhdr{Different Feature Sets}
Having constructed a successful predictive model, we now ask, ``What features are most valuable when modeling consumer choice?''
In the following set of experiments, we contrast the performance of SVM models
trained on different sets of graph and metadata features in order to better
understand how consumers make purchasing decisions.

When doing prediction, are network features as valuable as metadata features such as product price and seller rating?
We first compare the performance of a SVM trained on network (graph) features
versus a SVM trained on metadata (seller profile and product description) features. We
find that the P@1 of the graph features SVM is only slightly worse than the
Full model, whereas the metadata features SVM is not much
better than Random, as displayed in Table~\ref{table:predResults}.
Most notably, there is a large performance gap between the graph features SVM compared to the metadata
features SVM, as illustrated in Figures~\ref{fig:prediction}(d),(h).  This
implies that prediction using only seller and product information is
inadequate, the social and trade networks in the neighborhood of the buyer and
seller must be taken into account. Buyers likely do not just use seller profile and product information when making purchasing decisions.

Given that network features are essential when predicting consumer choice, what
type of network features are more valuable for prediction: direct features or
indirect features (i.e., clustering coefficients, PageRanks)? For this experiment, we
train SVM models on direct and indirect network features, the different graph
feature classes we use are listed in Table~\ref{table:featureset}. Note that when comparing direct and indirect network
features, we include all metadata features in both sets.
Figures~\ref{fig:prediction}(c),(g) illustrate the large
performance gap between the \textit{Meta + Direct} SVM vs the \textit{Meta +
  Indirect} SVM, labeled as \textit{Direct} and
\textit{Indirect} respectively in the figures. We observe that direct
graph information provided by buyer-seller edges is significantly more valuable than the collective
information provided by other edges in the local neighborhoods. This is
not surprising because powerful information content is present in direct
edges. Historical buyer-seller message volume can be indicative of an existing social
relationship or historical product queries, while historical
buyer-seller trade volume can be indicative of customer loyalty and trust with the seller.

Which network (contact, message, trade) is most useful for predicting consumer choice?
In this section, we contrast SVM models trained on each of the separate Taobao
networks. We include all metadata features with each set of network features in this experiment as well,
prediction results are displayed in Table~\ref{table:predResults}.
The performance of the 3 network feature sets versus the number of sellers in the cluster are displayed in Figures~\ref{fig:prediction}(b),(f).
Our experiment demonstrates that the message network is the most valuable
network to utilize when predicting consumer choice. One possible explanation for
this finding is that historical message volume is an indicator of familiarity between buyer and
seller, i.e., an existing trust relationship between buyer and seller.
Historical message volume can also indicate previous potential purchase interest.

We also observe that prediction with the trade network is slightly better than that with the
contact network. This suggests that customer loyalty, the primary
trade network feature we use, is a more important indicator of consumer choice
than the network of contacts. Our explanation for this is that, inherently, contact links are less
valuable than trade or message links in such social networks. It takes little
effort to add someone to a friends list, as it is a one-time
operation. In contrast, maintaining a conversation requires an investment of time and
mutual interest on the part of both parties. Forming a trade link is arguably
the most costly as it requires currency and an actual transaction to make the connection.

\xhdr{Per-Category Performance}
After performing prediction with a single SVM ranking model, the next question
to ask is, ``Can we perform better prediction through the use of multiple
models?'' To answer this question, we segment all historical transactions in our dataset into their respective product categories, and train
separate SVM ranking models for each category. The results of our model testing
are displayed as \textit{Aggregated} in Figure~\ref{fig:prediction}(a),(e).
As expected, the aggregated performance of the category SVMs is slightly
better than the single Full SVM, with a P@1 of 0.58 compared to 0.56.

Our study of consumer choice prediction demonstrates that in social commerce sites such as Taobao,
user communication and social activity is the primary influence upon consumer
choice. Utilizing primarily social networking features, we are able to construct
an SVM model that can predict, for the case of a buyer choosing from among 10
possible sellers, the correct seller 42\% of the time, approximately 4 times
better than random. When faced with a selection of substitute goods offered by different
sellers, buyers will not just choose their preferred seller through simple
heuristics regarding price or rating. We can imagine that buyers utilize many sources of
information (seller history, advice of friends, seller's messages), and each buyer
processes the information in their own way in order to make a personal purchasing
decision. Although we cannot say with certainty what buyers are thinking, we can definitively state that the social graph in
which the buyer and seller are embedded is the best feature to look at when
predicting consumer choice.

\section{Conclusion}
\label{sec:conclusion}
Our work analyzes the activities of one million users of the Chinese social
commerce site Taobao. Through the study of directed closure rules, we
empirically verify that implicit information passing is present in the Taobao
network, and show that communication between buyers is a fundamental driver
of purchasing activity. We then investigate the directed triadic closure process
and explain how link formation is highly dependent upon the distribution of buyer/seller
roles for the nodes of a social commerce network.
Third, we use Taobao review data to demonstrate how high seller ratings are associated
with product price premiums, and thus quantify a price for trust.
Finally, we develop a machine learning model to accurately predict consumer choice, and
demonstrate that the social network is the most important feature in predicting
how consumers choose their transaction partners.

We hope that our study will motivate future research into social shopping, as
well as give impetus to established e-commerce companies to add more social networking
features. Future areas of related study include: analysis of user browsing
data to develop refined consumer choice models for social commerce,
study of information passing while factoring in both buyer-buyer and
buyer-seller trust relationships, and viral marketing to influence consumer choice
in social commerce.

\xhdr{Acknowledgements} Research
was in-part supported by NSF
CNS-1010921,  
NSF IIS-1016909,    
Albert Yu \& Mary Bechmann Foundation, IBM,
Lightspeed, Microsoft and Yahoo.

\bibliographystyle{abbrv}

\end{document}